# Detailed abundance determination of metal-poor stars with X-Shooter – I. Unusual chemistry in halo stars

Benjamin D. C. Lowe ,[1★] Thomas Nordlander,[2] Luca Casagrande ,[1] Gary S. Da Costa ,[1] Norbert Christlieb,[3] Sarah E. Aquilina,[1] Tomasz Różański [1] and Giacomo Cordoni [1]

[1] *Research School of Astronomy and Astrophysics, Australian National University, Canberra, ACT 2611, Australia*
[2] *Theoretical Astrophysics, Department of Physics and Astronomy, Uppsala University, Box 516, SE-751 20 Uppsala, Sweden*
[3] *Zentrum für Astronomie der Universität Heidelberg, Landessternwarte, Königstuhl 12, D-69117 Heidelberg, Germany*

Accepted 2026 January 21. Received 2026 January 9; in original form 2025 October 28

## ABSTRACT

We present a detailed chemical analysis study of 16 candidate metal-poor stars, previously identified with 2dF+AAOmega, using X-Shooter spectra and the Korg 1D local thermodynamic equilibrium spectral synthesis code. We confirm the earlier metallicity estimates and reveal six extremely metal-poor ([Fe/H] $< -3$) stars in the current sample. Two of these stars, including the most metal poor at [Fe/H] $= -3.89 \pm 0.07$, are kinematically associated with the *Gaia*–Sausage–Enceladus (GSE) accretion event, increasing the number of known GSE stars with [Fe/H] $< -3.5$ to eight. From the X-Shooter spectra we determine abundances for 16 elements, with the element-to-iron abundance ratios generally consistent with high-resolution studies of Galactic halo stars. Within the sample, we identify three peculiar stars: the first is a GSE nitrogen enhanced metal-poor ([N/Fe] $= 1.60 \pm 0.10$ and [C/Fe] $= 0.23 \pm 0.08$) star with unusually high Na ([Na/Fe] $= 2.26 \pm 0.07$) and Li ($A(\mathrm{Li})_{\mathrm{3DNLTE}} = 1.90 \pm 0.08$) abundances, but lacking enhancements in [Al/Fe] or [Mg/Fe]. The second is a halo r-II star significantly enhanced in Sr ([Sr/Ba] $= 0.39 \pm 0.08$), suggesting mixture of r-process and s-process enrichment, uncommon for r-II stars. Whilst the third is a halo star very depleted in N ([N/Fe] $< -1.11$), with low C ([C/Fe] $= -0.33 \pm 0.08$) and otherwise 'normal' [X/Fe] abundances, suggesting enrichment with Type II supernova that proceeds enrichment from massive asymptotic giant branch stars. This study reveals the substantial degree of chemical diversity in the stellar populations which assembled the early Milky Way.

## 1 INTRODUCTION

Some of the oldest stellar objects observable in the Universe today are the extremely metal-poor (EMP) stars with [Fe/H] $< -3.0$ [1] (T. C. Beers & N. Christlieb 2005). These elusive objects have witnessed at least 10 billion years of history unfold, with the most metal poor of these EMPs being presumably formed out of the gas enriched by the supernovae from the very first stars: the metal-free Population III (Pop. III) stars (e.g. R. S. Klessen & S. C. Glover 2023). This hypothesized class of stars has not yet been detected directly, despite many efforts undertaken to find them directly at high redshift (e.g. S. P. Oh, Z. Haiman & M. J. Rees 2001; E. Scannapieco, R. Schneider & A. Ferrara 2003; T. H. Greif et al. 2009; E. Zackrisson et al. 2011, 2012; C.-E. Rydberg et al. 2013; L. Mas-Ribas, M. Dijkstra & J. E. Forero-Romero 2016; S. Riaz, T. Hartwig & M. A. Latif 2022), though several studies with the *James Web Space Telescope* (*JWST*) have indicated potential Pop. III signatures (e.g. S. Cai et al. 2025; D. Ďurovčíková et al. 2025; S. Fujimoto et al. 2025; C. Mondal et al. 2025; T. Morishita et al. 2025).

Because of this difficulty, efforts have instead focused on studying EMP stars in our Galaxy, particularly those that are the most iron deficient. These stars contain the best preservation of Pop. III progenitor signatures that we can detect, with many great advances already made at studying these lowest metallicity stars (e.g. M. S. Bessell & J. Norris 1984; N. Christlieb et al. 2002; A. Frebel et al. 2005; E. Caffau et al. 2011b; D. S. Aguado et al. 2018; E. Starkenburg et al. 2018).

One such star, SMSS J031300.36−670839.3, discovered using the SkyMapper Southern Sky (SMSS) survey (C. Wolf et al. 2018; C. A. Onken et al. 2019, 2024), is the most iron-poor star found, having an upper limit of [Fe/H] $< -7.52$ (S. C. Keller et al. 2014; M. S. Bessell et al. 2015). The star shows large overabundances of carbon, oxygen, and magnesium relative to calcium. These abundance patterns are consistent with this star being produced from a Pop. III progenitor with a mass in the range 10–60 $\mathrm{M}_\odot$ at modest explosion energies (T. Nordlander et al. 2017).

Another star, SMSS J160540.18−144323.1, has the lowest detected Fe abundance with [Fe/H] $= -6.2 \pm 0.2$ (T. Nordlander et al. 2019). Interestingly, this star is also strongly C-enhanced, having [C/Fe] = $3.9 \pm 0.2$, but otherwise has a uniform abundance trend across its measured elements. This level of enhancement seen only in C, with otherwise normal abundances in Mg, Ca, and Ti, is attributed to a low-mass Pop. III progenitor ($\approx 10\,M_\odot$) fallback supernovae exploding at low energies. Progenitors with masses greater than $20\,M_\odot$ fail to reproduce these abundance patterns. Despite also having large C abundances, the star SMSS J031300.36−670839.3 discussed earlier shows enhancements in Mg and O not seen in SMSS J160540.18−144323.1, making the fallback Pop. III supernovae scenario for it infeasible.

★ E-mail: ben.lowe@anu.edu.au
[1] $[\mathrm{X/H}] = \log(N_\mathrm{X}/N_\mathrm{H})_\star - \log(N_\mathrm{X}/N_\mathrm{H})_\odot$, where $N_\mathrm{X}$ is the number density for element X.

The most pristine metal-poor star observed to date is SDSS J0715−7334 from A. P. Ji et al. (2025), a Large Magellanic Cloud (LMC) star which at [Fe/H] = $-4.53 \pm 0.20$, has a low C upper-limit of [C/Fe] $< -0.22$, very rare for stars this metal poor. The total metallicity of this star ($Z < 7.8 \times 10^{-7}$) is lowest seen in the literature, surpassing the previously lowest overall metallicity star (E. Caffau et al. 2011a). This remains consistent with the independent analysis of G. Limberg et al. (2025), even when accounting for lower signal-to-noise (SNR) spectra. Therefore, this star is substantially more metal poor than the earliest galaxies seen by *JWST* at high redshifts (e.g. S. Fujimoto et al. 2025; T. Morishita et al. 2025; K. Nakajima et al. 2025). Star SDSS J07157334 will be vital in providing constraints on the earliest stages of the Universe, particularly with understanding the Pop. III stars.

A handful of individual stars like these have enhanced our understanding of the early Universe. Many studies utilizing large-scale survey missions have therefore sought to search the night sky for these elusive stars (e.g. N. Christlieb et al. 2008; T. Schörck et al. 2009; A. Frebel 2010; S. Salvadori et al. 2010; E. Caffau et al. 2013; L. M. Howes et al. 2015, 2016; G. S. Da Costa et al. 2019; A. Arentsen et al. 2020; M. N. Ishigaki et al. 2021; T. S. Li et al. 2022; X. Hou, G. Zhao & H. Li 2024; B. D. C. Lowe et al. 2025). Upcoming large-scale surveys like 4MOST and WEAVE (e.g. R. S. Jong et al. 2019; S. Jin et al. 2024) will help to further expand the pool of known EMPs.

Understanding the chemical makeup of metal-poor stars extends beyond trying to learn about Pop. III stars, as they are also vital in providing observational constraints on the formation and evolution of our Galaxy. For the Galactic disc, EMP studies have begun providing a new perspective on understanding how the disc formed. The star SDSS J102915+172927 (E. Caffau et al. 2011b, 2012) with [Fe/H] = −4.73, is confined to a prograde disc orbit with $z_{\rm max} < 3$ kpc and $e = 0.12 \pm 0.01$ (F. Sestito et al. 2019). Unlike other stars at similar metallicities, this star has little C-enhancement, having an upper-limit of [C/Fe] $< 0.6$ from 3D hydrodynamic model atmospheres (C. Lagae et al. 2023). It also has low Al and Na abundances, alongside a very low Li abundance (E. Caffau et al. 2024), suggesting formation through dust cooling (R. S. Klessen & S. C. Glover 2023).

Another prograde EMP disc star, P1836849 with [Fe/H] = $-3.3 \pm 0.1$, has low $\alpha$-element abundances, but high abundances of Cr and Mn. A. Dovgal et al. (2024) has suggested that these abundances can be reproduced by a Pop. III progenitor with masses $10\,M_\odot$ or $17\,M_\odot$. It has been shown through simulations that galactic discs form via accretion, yielding metal-poor stars on both prograde and retrograde orbits (e.g. I. B. Santistevan et al. 2021).

Many studies are now focusing on the disc to prove this observationally by expanding the pool of disc EMP stars to confirm if this is true or not (e.g. F. Sestito et al. 2019, 2020; A. Chiti et al. 2021; G. Cordoni et al. 2021; E. Fernández-Alvar et al. 2021; C. L. Kielty et al. 2021; M. Bellazzini et al. 2024; B. D. C. Lowe et al. 2025).

The Galactic halo, known for its chemical abundance inhomogeneity and accretion origins, has a wide range of EMP stars that provides valuable insights into the early history of the Galaxy. For example, the *Gaia*–Sausage–Enceladus (GSE; V. Belokurov et al. 2018; A. Helmi et al. 2018) accreted substructure in the halo is the debris of the last major merger experienced by the Galaxy about 8–11 billion years ago (e.g. F. Vincenzo et al. 2019; V. Belokurov et al. 2020). Finding and analysing metal-poor GSE stars will allow its history to be constrained.

One such analysis was performed on the metal-poor GSE star LAMOST J0804+5740 with [Fe/H] = −2.38, revealing it to be r-process enhanced, with [Eu/Fe] = 0.80 (Y. Lin et al. 2025). This is also the first GSE star with extreme enhancements in actinide elements like Th and U relative to r-process elements (otherwise known as an actinide-boost star). Other metal-poor studies have revealed that the metallicity distribution function for the GSE peaks at [Fe/H] $\approx -1.2$ with a metal-weak tail extending to [Fe/H] $\approx -3.0$ (e.g. D. K. Feuillet et al. 2020; R. P. Naidu et al. 2020; P. Bonifacio et al. 2021; G. Limberg et al. 2022). Despite this, a small handful of GSE candidates are genuine EMP stars (e.g. G. Cordoni et al. 2021; D. Yong et al. 2021a; R. Zhang et al. 2024; V. M. Placco et al. 2025). These studies all showed that the fraction of metal-poor stars in the GSE is lower compared to the Galaxy, suggesting that the large number of EMP stars in the halo was contributed not by massive dwarf galaxies like the GSE progenitor, but rather by accreting smaller, ultrafaint dwarf galaxies.

Metal-poor studies in the literature have revealed the importance of studying these stars across the Galaxy, helping constrain both the progenitor Pop. III stars, and the formation history of our Galaxy. Though the relative lack of them, particularly in the GSE and Galactic disc, hinders this significantly. In response to this, we are leading a survey on the multifibre 2dF instrument coupled with the AAOmega spectrograph (W. Saunders et al. 2004; R. Sharp et al. 2006) to target metal-poor candidates across the disc (B. D. C. Lowe et al. 2025). Here, we provide a detailed chemical characterization of 16 previously identified metal-poor stars (six of which are EMPs) across the halo, prograde disc and GSE using the X-Shooter medium-resolution spectrograph (J. Vernet et al. 2011). In what follows we present our observations and data processing (Section 2), chemical abundance derivation (Section 3), our results (Section 4) and then the discussion (Section 5).

## 2 OBSERVATIONS AND DATA PROCESSING

### 2.1 Target Selection, observations, and data reduction

The sample of 16 metal-poor stars in this work was identified from the *Gaia* BP/RP and 2dF+AAOmega study by B. D. C. Lowe et al. (2025). Five of the stars were identified as EMP stars, with the remaining 11 chosen based on their kinematic classifications. The magnitudes and *Gaia* DR3 coordinates are found in Table 1. The remaining EMP candidates from B. D. C. Lowe et al. (2025) will be presented in a forthcoming work (B. D. C. Lowe et al., in preparation).

Observations of the 16 metal-poor stars were performed in service mode from April to July 2024 (Programme ID: 113.26N5.001), with the high-efficiency spectrograph X-Shooter (J. Vernet et al. 2011) on Unit Telescope 2 (UT2, Kueyen) of the Very Large Telescope (VLT) at Cerro Paranal Observatory. Our observations with X-Shooter utilized the Ultraviolet-Blue (UVB; 3000–5595 Å) and Visible (VIS; 5595–10240 Å) arms, with a slit width and length of 1.0 arcsec × 11 arcsec and 0.9 arcsec ×

**Table 1.** List of programme stars studied. Their *Gaia* DR3 source ID, RA and Dec. coordinates (in the ICRS reference frame at the J2016.0 epoch), galactic coordinates (galactic longitude, $l$, and galactic latitude, $b$), parallax $\pi$, apparent magnitudes $m_{\rm G}$, reddening $E(B-V)$[from D. J. Schlegel, D. P. Finkbeiner & M. Davis (1998), rescaled as per L. Casagrande et al. (2019)] and kinematic groupings from B. D. C. Lowe et al. (2025) are provided.

| Star ID | *Gaia* DR3 ID | RA | Dec | $l$ (deg) | $b$ (deg) | $\pi$ (mas) | $m_{\rm G}$ (mag) | $E(B-V)$ (mag) | Orbit |
|---|---|---|---|---|---|---|---|---|---|
| ra_0103-7050_s163 | 4689369645972422784 | 00:53:36.23 | −71:04:47.35 | 302.7 | −46.0 | 0.74 ± 0.03 | 15.6 | 0.04 | Prograde disc |
| ra_0834-5220_s316 | 5321186578181666176 | 08:36:16.70 | −52:56:35.63 | 269.9 | −7.3 | 0.65 ± 0.05 | 16.8 | 0.51 | Prograde disc |
| ra_1604-2712_s24 | 6043161513972081024 | 16:01:40.83 | −26:40:51.20 | 347.3 | 19.4 | −0.02 ± 0.04 | 15.5 | 0.10 | Halo |
| ra_1604-2712_s292 | 6042817710432143104 | 16:02:07.43 | −27:42:03.79 | 346.6 | 18.6 | 0.19 ± 0.06 | 16.8 | 0.12 | Halo |
| ra_1624-2150_s278 | 6052240868671385472 | 16:24:12.43 | −21:19:27.92 | 355.2 | 19.3 | 0.29 ± 0.08 | 17.3 | 0.43 | Halo |
| ra_1633-2814_s130 | 6044035900595959424 | 16:34:49.20 | −28:28:09.76 | 351.2 | 12.8 | 0.20 ± 0.08 | 16.9 | 0.50 | GSE |
| ra_1633-2814_s284 | 6044482989511923584 | 16:33:14.64 | −28:03:05.13 | 351.3 | 13.4 | 0.25 ± 0.10 | 17.1 | 0.40 | GSE |
| ra_1648-0653_s38 | 4340844491685350912 | 16:45:16.01 | −06:40:41.58 | 11.0 | 24.2 | 0.25 ± 0.08 | 17.0 | 0.41 | Prograde disc |
| ra_1656-1433_s143 | 4140336967130944640 | 16:56:13.33 | −14:06:46.88 | 6.1 | 17.7 | 0.05 ± 0.03 | 15.4 | 0.67 | Halo |
| ra_1658-2454_s22 | 4113281490693917568 | 16:54:43.46 | −24:29:20.57 | 357.2 | 11.9 | 0.13 ± 0.03 | 15.3 | 0.32 | Halo |
| ra_1659-2154_s114 | 4127849985388551808 | 17:00:18.31 | −21:05:57.14 | 0.8 | 12.9 | 0.03 ± 0.09 | 17.2 | 0.22 | Halo |
| ra_1709-2130_s102 | 4127737388534814848 | 17:08:50.26 | −20:44:38.97 | 2.3 | 11.5 | 0.14 ± 0.05 | 16.0 | 0.29 | Halo |
| ra_1752-4300_s214 | 5956402620533315200 | 17:50:34.43 | −43:13:25.11 | 348.2 | −8.2 | 0.03 ± 0.09 | 16.1 | 0.14 | Halo |
| ra_1752-4300_s269 | 5956280128069527808 | 17:54:29.31 | −43:00:13.62 | 348.7 | −8.7 | 0.11 ± 0.06 | 16.5 | 0.10 | Halo |
| ra_1752-4300_s6 | 5957174928716116224 | 17:48:06.46 | −42:50:08.67 | 348.3 | −7.6 | 0.04 ± 0.04 | 15.5 | 0.20 | Halo |
| ra_1853-3255_s45 | 6735735401460422528 | 18:52:30.98 | −33:38:05.23 | 2.4 | −14.8 | 0.14 ± 0.03 | 14.9 | 0.02 | Halo |

*Note.* The full table is available in machine-readable format in the electronic version of the paper.

11 arcsec, yielding resolving powers of $R = 5400$ and 8900, respectively.

### 2.2 Data normalization and processing

Continuum normalization was performed using SUPPNET (T. Różański et al. 2022), a fully convolutional neural network trained on a diverse set of synthetic and empirically normalized high-resolution spectra. The method operates directly on order-merged data and outputs a predicted pseudo-continuum. This approach enables reproducible normalization of large batches of spectra with high precision, achieving a root mean square normalization accuracy better than ∼1 per cent, even for spectra affected by moderate noise, rotational broadening, and emission-line features. The automated and deterministic nature of SUPPNET significantly reduces both the subjectivity and effort associated with manual normalization methods, ensuring consistency and enabling straightforward cosmic ray removal and averaging across multiple spectra.

Spectra were stacked using our custom-built data processing code. Given that several of our stars required multiple observations to reach desired signal-to-noise (S/N) in both the UVB and VIS arms, the exposures for these stars are spread across multiple nights. Therefore, once the data were normalized, for stars with multiple exposures, we aligned spectra by manually shifting to known strong absorption lines in the exposure with the highest S/N: for UVB, we shifted to $\rm H_\beta$ (4861.35 Å), and for VIS, we shifted to $\rm H_\alpha$ (6562.79 Å). The combined spectra were then shifted to rest using the radial velocities described at the end of the section.

Cosmic rays were then removed for each observation by taking $3\sigma$ of the noise (tested by manual inspection), and removing pixels above this threshold (which we assumed to be cosmic rays). We also rejected ±1 pixels either side of the cosmic ray source pixels. A weighted average (using the S/N of each spectrum) of the observations was taken to generate the stacked normalized spectra.

Heliocentric-corrected radial velocities (alongside barycentric-corrected velocities) were measured separately for both the UVB and VIS arms of the stacked spectra: for UVB, we fitted Voigt functions to H $\beta$, H $\gamma$ (4340.47 Å) and H $\delta$ (4101.73 Å). Whilst for VIS, we fitted Voigt functions to H $\alpha$ and the Ca II triplet lines (8498.23, 8542.31, and 8662.36 Å). Weighted averages using the radial velocity errors as weights of each arm were taken, with the uncertainties from the weighted standard error of the mean. These can be found in Table 2. Wavelength calibrations issues have been known to impact radial velocity measurements across the two arms for X-Shooter spectra, with 4.477 and 1.001 km s$^{-1}$ offsets necessary for the UVB and VIS arms, respectively (e.g. H. Sana et al. 2024). These have been applied to our measurements. For this study, we adopted the radial velocity as the weighted average between the two spectral arms. Since the VIS measurements have smaller associated errors, they have a greater influence on the combined radial velocity value.

## 3 ANALYSIS

### 3.1 Stellar parameters

The effective temperature ($T_{\rm eff}$) and surface gravity ($\log g$) for the stars in the current sample were derived in B. D. C. Lowe et al.

**Table 2.** Observation log for observed stars. The number of observations, exposure times for each observation, alongside the average radial velocities and S/N for both the UVB and VIS arms are provided. The radial velocity is from the stacked spectra. The S/N listed are the combined values for each arm (added in quadrature).

| Star ID | $N_{exp}$ | Exp (s) | $S/N_{UVB}$ | $S/N_{VIS}$ | $RV_{helio,avg}$ (km $s^{-1}$) |
|---|---|---|---|---|---|
| ra_0103-7050_s163 | 1 | 1800 | 125.8 | 129.6 | $-17.7 \pm 0.2$ |
| ra_0834-5220_s316 | 2 | 2760 | 88.7 | 145.3 | $136.8 \pm 0.2$ |
| ra_1604-2712_s24 | 1 | 2700 | 152.5 | 207.4 | $171.5 \pm 0.1$ |
| ra_1604-2712_s292 | 2 | 2100 | 84.4 | 95.3 | $15.4 \pm 0.3$ |
| ra_1624-2150_s278 | 3 | 2520 | 108.1 | 149.7 | $223.8 \pm 0.3$ |
| ra_1633-2814_s130 | 4 | 3120 | 114.5 | 237.3 | $174.5 \pm 0.2$ |
| ra_1633-2814_s284 | 4 | 3120 | 122.5 | 209.1 | $162.4 \pm 0.2$ |
| ra_1648-0653_s38 | 3 | 3120 | 126.6 | 184.8 | $-36.0 \pm 0.2$ |
| ra_1656-1433_s143 | 2 | 3120 | 145.0 | 347.5 | $-76.9 \pm 0.1$ |
| ra_1658-2454_s22 | 1 | 2820 | 130.7 | 227.9 | $39.3 \pm 0.1$ |
| ra_1659-2154_s114 | 4 | 3120 | 133.4 | 211.1 | $-140.4 \pm 0.1$ |
| ra_1709-2130_s102 | 2 | 2520 | 127.2 | 214.5 | $280.5 \pm 0.1$ |
| ra_1752-4300_s214 | 2 | 2520 | 132.7 | 202.5 | $-163.6 \pm 0.1$ |
| ra_1752-4300_s269 | 2 | 3120 | 131.5 | 190.0 | $3.0 \pm 0.1$ |
| ra_1752-4300_s6 | 1 | 2060 | 129.1 | 207.4 | $322.4 \pm 0.1$ |
| ra_1853-3255_s45 | 1 | 1860 | 185.6 | 215.4 | $-146.5 \pm 0.1$ |

*Note.* The full table is available in machine-readable format in the electronic version of the paper.

(2025). In particular, $T_{\rm eff}$ was found using the COLTE[2] code: estimating $T_{\rm eff}$ by applying calibrated photometric colour–$T_{\rm eff}$ relations of L. Casagrande et al. (2021). For $\log g$, this was derived from either the absolute bolometric magnitude, $T_{\rm eff}$ and stellar mass (if $\pi \geq 3\sigma$), or from isochrones (if $\pi < 3\sigma$). The microturbulence velocity ($v_{\rm mic}$) for each star was determined using equation (4) from S. Buder et al. (2025), which resulted in values between 1.25 and 2.21 km $s^{-1}$. Uncertainties in $v_{\rm mic}$ were derived from error propagation using the errors in $T_{\rm eff}$ and $\log g$. We assumed that these metal-poor stars have long since spun-down to velocities below the instrumental resolution of X-Shooter, and thus we set the projected rotational velocity $v \sin i = 0\,{\rm km\,s^{-1}}$.

### 3.2 Spectroscopic analysis

Metallicities and stellar abundances were derived using the spectral synthesis code KORG (A. J. Wheeler et al. 2022, 2023), assuming local thermodynamic equilibrium (LTE) and one-dimensional geometry. We used the Vienna Atomic Line Database (N. E. Piskunov et al. 1995; F. Kupka et al. 1999; T. Ryabchikova et al. 2015) linelist, alongside the MARCS model atmospheres (B. Gustafsson et al. 2008) within KORG.

For this work, we employed KORG's FIT_SPECTRUM method, which uses $\chi^2$ minimization to fit a synthetic spectra to the given observed data in a line by line analysis. This requires the input of the observed flux, flux errors, and wavelengths (in vacuum), alongside the desired line list, wavelength fitting windows, spectral resolution, fixed stellar parameters and initial guess of the abundances. In most cases, we used ADJUST_CONTINUUM in the FIT_SPECTRUM method to adjust the continuum with the best-fitting linear correction to match our observed continuum.

For each element, abundances were determined by running the first iteration using the initial guesses, then running it again using the previous fitted abundance. Once all elements were measured, abundances were re-determined in FIT_SPECTRUM, with all of the other measured abundances set as fixed parameters. The procedure was carried out for nine iterations, by which point the difference between successive steps had converged to < 0.01 dex. Since our stars are metal poor, we set $[\alpha/{\rm Fe}] = 0.4$ for the first iteration, then each subsequent iteration used the weighted average of the $\alpha$-elements (Mg, Si, Ca, and Ti). For elements we did not measure, like O, their abundance was set by the $[\alpha/{\rm Fe}]$ value. KORG has no inbuilt region masking within the provided fitting windows. Therefore to mask-out lines impacting continuum placement, we inflated the errors of the undesired pixels to a very large value ($10^{10}$), so that they are effectively ignored by KORG. Below, we discuss how we used KORG to measure our metallicities and chemical abundances.

#### 3.2.1 *Metallicities*

To ensure we measured reliable and accurate metallicities,[3] we used the Fe I lines employed by E. Caffau et al. (2013, see their table 3), a previous X-Shooter study of metal-poor stars. For the first iteration, we set the initial guess to be the 2dF+AAOmega [Fe/H] value, before taking the fitted value for the second iteration. This process was repeated once all elements were measured, setting them as fixed parameters whilst adopting the previously determined [Fe/H] value as our initial guess. The adopted metallicity value is the weighted average of all measured Fe I lines.

#### 3.2.2 *Abundances*

For our chemical abundances linelist, we first generated synthetic equivalent widths (EW) for each feature (using as reference $T_{\rm eff} = 4500$ K, $\log g = 1.5$, $[{\rm Fe/H}] = -3.0$, and $[\alpha/{\rm Fe}] = 0.4$), then visually checked what EW can be seen in the observed data set. For the blue, we set the limiting EW at 80 mÅ, and for the red, we set the limiting EW at 60 mÅ. The same list of lines was used for all stars in our sample, making our analysis nearly differential. Window regions were selected to cover each line with a width of at least ±3 Å, available in Tables S1 and S2 (for Na I – see bottom of section) of Supporting Information.

[2]https://github.com/casaluca/colte

[3]In this work, we take metallicity to be [Fe/H].

Using the metallicities derived as described in Section 3.2.1, we computed abundances using atomic lines of the following species: Na I, Mg I, Al I, Si I, Ca II, Sc II, Ti II, Cr I, Mn I, Co I, Ni I, Sr II, Ba II, and Eu II. We also computed abundances using molecular bands of CH (~4300 Å) and NH (~3360 Å). The abundances of O and other $\alpha$-elements that could not be measured from these spectra was set by the [$\alpha$/Fe] abundance, which in our case is [O/Fe] = 0.4. For the first iteration, we used [X/Fe] = 0 as the initial guess, then adopted the fitted value as the guess for the second iteration. Like with metallicity, this process was repeated once all the elements were measured, remeasuring our specified element whilst setting the previous abundances as fixed parameters. The initial guess was set by its previous value. For elements with multiple lines, a weighted average was performed at each iteration, and then used as the abundance value for the element. These were then used as fixed parameters when redetermining abundances.

For Ca II, the strong absorption features of the 8498.23, 8542.31, and 8662.36 Å lines caused difficulties with KORG's automatic continuum adjustment feature. The reason for this remains unknown, but to fix this, we turned off the continuum adjuster and manually adjusted the synthetic continuum to match the observed continuum. We also applied non-local thermodynamic equilibrium (NLTE) corrections to the fitted abundance, adopting the corrections from Y. Osorio et al. (2022). These were interpolated using a piecewise linear interpolator across the stellar parameter grids ($T_{\rm eff}$, $\log g$, and [Fe/H]) for each measured Ca II line. Typical NLTE corrections span between −0.20 and −0.40 dex, and result in abundances that are lower than the LTE values. The NLTE corrections and continuum offsets for the Ca II triplet lines are available in Table S3 of Supporting Information.

Li I at 6707.81 Å was measured in only one spectrum from our sample. Given the strong 3D NLTE effects present, KORG was used to measure the LTE abundances, then the Li abundance predictor code BREIDABLIK[4] (E. X. Wang et al. 2021, 2024) was used to perform the corrections ($\Delta_{\rm 3DNTLE} = -0.08$).

We applied evolutionary mixing corrections to our fitted C abundances using V. M. Placco et al. (2014)[5] for our stars. These range from +0.00 dex for our dwarfs, to +0.46 dex higher than the measured values for our coolest, lowest $\log g$ giants. The evolutionary corrections for C are available in Table S4 of Supporting Information. [N/Fe] was not evolutionary corrected in this work, though we would expect a large [C/Fe] correction would correspond to a corrected [N/Fe] value being lower than its observed value.

For Na I, the presence of interstellar absorption close to the stellar lines required us to define star-specific windows to avoid contamination (available in Table S2 of Supporting Information). For three stars, the interstellar lines crossed over with the stellar lines, forcing us to discard them completely.

### 3.3 Error analysis and upper limits

We estimated the uncertainty in our measured metallicities and chemical abundances by perturbing the stellar parameters $T_{\rm eff}$ by +100 K, $\log g$ by +0.3, and $v_{\rm mic}$ by +0.3 km s$^{-1}$, then adopting these values separately in KORG. The difference in the inferred abundances compared to the reference value was taken as the error. These were then added in quadrature to get the total error for the measurement. The median errors for our abundances for the perturbations are given in Table 3.

[4]https://github.com/ellawang44/Breidablik
[5]https://vplacco.pythonanywhere.com/

**Table 3.** Median errors for our chemical abundances and metallicities ordered by atomic number. Errors are split between those from perturbed $T_{\rm eff}$, $\log g$, and $v_{\rm mic}$.

| Element | Median error ($T_{\rm eff}$ + 100 K) | Median error ($\log g$ + 0.3) | Median error ($v_{\rm mic}$ + 0.3 km s$^{-1}$) |
|---|---|---|---|
| C (CH) | 0.23 | 0.11 | 0.00 |
| N (NH) | 0.24 | 0.12 | 0.03 |
| Na I | 0.12 | 0.04 | 0.11 |
| Mg I | 0.10 | 0.07 | 0.09 |
| Al I | 0.07 | 0.01 | 0.10 |
| Si I | 0.15 | 0.12 | 0.10 |
| Ca II | 0.06 | 0.01 | 0.02 |
| Sc II | 0.04 | 0.14 | 0.07 |
| Ti II | 0.04 | 0.09 | 0.09 |
| Cr I | 0.12 | 0.00 | 0.04 |
| Mn I | 0.17 | 0.04 | 0.03 |
| Fe I | 0.11 | 0.08 | 0.13 |
| Co I | 0.13 | 0.02 | 0.03 |
| Ni I | 0.08 | 0.01 | 0.03 |
| Sr II | 0.07 | 0.08 | 0.18 |
| Ba II | 0.07 | 0.09 | 0.02 |
| Eu II | 0.16 | 0.11 | 0.05 |

*Note.* The full table is available in machine-readable format in the electronic version of the paper.

Upper-limits were determined differently for our atomic and molecular features. For the atomic lines, we used equation (6') from R. Cayrel (1988), which for a given S/N, spectral resolution and pixel steps, provides the minimum observed EW for the given spectral quality. We took 3$\sigma$ of this to be our detection limit. This was done for the strongest line belonging to the specific element in the spectral window. To convert this into abundance space, we used KORG's EWS_TO_ABUNDANCES method, which uses the linear part of the curve-of-growth to perform the conversion for a given model atmosphere. Synthetic spectra were generated at this abundance using KORG's SYNTHESIZE, and then visually compared to the observed spectrum. If the observed measured abundance was lower than the minimum, then the line was labelled as non-detected, with the upper-limit being the minimum abundance.

If an element had several upper limits from different lines, we chose the smallest one. If at least one line had a detection, then the element was flagged as detected, only adopting the abundance of the detected line(s).

For the molecular lines, we generated synthetic spectra from [X/Fe] = −4.0 to +4.0 in steps of 0.01 dex, then calculated the $\chi^2$ for each. This process was repeated for an 'empty spectrum' at [X/Fe] = −10.0, using the assumption that the molecular feature was negligible below this point.

Upper limits were determined by comparing the same synthetic spectra to the 'empty spectrum'. After adopting the 3$\sigma$ level, a non-detection was given where the best-fitting spectrum had an observed abundance less than this threshold, adopting this as its upper-limit. For the detections, statistical fitting errors were taken from the average of the lower and upper error bar.

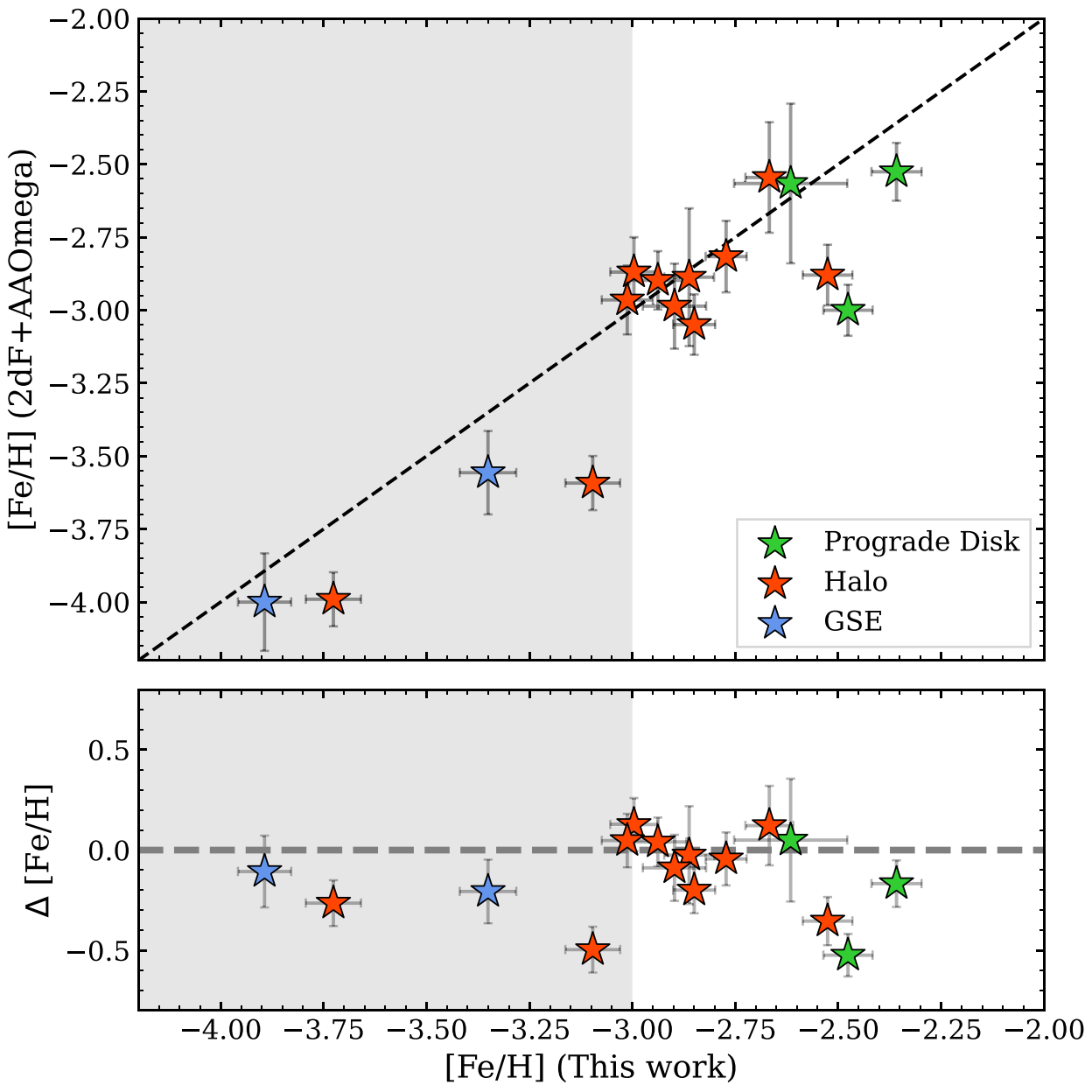

**Figure 1.** *Upper panels:* Metallicity comparison of the values from the X-Shooter spectra, against the 2dF+AAOmega values from B. D. C. Lowe et al. (2025). 1:1 line shown by diagonal dashed line. Shaded region refers to our measured metallicities with [Fe/H] $\leq -3.0$ . *Lower panels:* The difference (2dF+AAOmega minus X-Shooter) between metallicities derived from the X-Shooter and 2dF+AAOmega spectra; the dashed line is for zero difference.

## 4 RESULTS

### 4.1 Metallicities

We have analysed 16 metal-poor stars observed with the X-Shooter spectrograph. Metallicities were found to range from [Fe/H] $= -2.4$ to $-3.9$, with six stars having [Fe/H] $\leq -3.0$. As shown in Fig. 1, these line-by-line metallicity measurements correlate well with the metallicities derived from the 2dF+AAOmega spectra in B. D. C. Lowe et al. (2025). The mean Δ [Fe/H] (2dF+AAOmega minus X-Shooter) is $-0.13$ with a standard deviation of 0.20 dex. The values are consistent with similar studies (e.g. G. S. Da Costa et al. 2019; D. Yong et al. 2021a; W. S. Oh et al. 2023, 2024) that compared low- and high-resolution metallicities for the same stars. The X-shooter metallicities, alongside our derived $v_{\rm mic}$ (see Section 3.1) and the stellar parameters taken from B. D. C. Lowe et al. (2025) are listed in Table 4.

The weighted average radial velocities derived from this work (in Table 2) showed a mean difference and standard deviation of $0 \pm 10$ km s$^{-1}$ with the values given in B. D. C. Lowe et al. (2025). Given the small difference, we choose to adopt the orbital classifications defined in the 2dF+AAOmega analysis.

### 4.2 Abundance ratios

We show in Fig. 2 our chemical abundances for C (CH), N (NH), Na I, Mg I, Al I, Si I, Ca II, Sc II, Ti II, Cr I, Mn I, Co I, Ni I, Sr II, Ba II, and Eu II. All abundance measurements are 1D LTE, with Ca being NLTE-corrected. This ensures consistency with the literature comparison sample from D. Yong et al. (2013, 2021a, shown in grey), H. R. Jacobson et al. (2015), and A. F. Marino et al. (2019), with all of them adopting 1D LTE abundance measurements (except for Ca being NLTE-corrected). Points are separated into their kinematic groupings provided in Table 1, with upper-limits shown by the downward-facing arrows. The number of stars with measurements ($N$), alongside their mean ($\mu$) and standard deviation ($\sigma$) is given in each panel. Given the small sample sizes, there are no obvious differences between the different kinematic groups. We present our chemical abundances in Table 5, with our chemical abundance patterns shown in Fig. 3. Below we discuss each nucleosynthetic group in turn.

#### 4.2.1 *Light elements*

The light elements measured in this study include C and N. These elements are produced through stellar nucleosynthesis in evolved stars via the triple-alpha process and the CNO cycle, respectively.

**Table 4.** Stellar parameters for our sample with uncertainties. [Fe/H] and $v_{\rm mic}$ were derived in this work, whilst $T_{\rm eff}$ and $\log g$ (alongside their uncertainties) were taken from B. D. C. Lowe et al. (2025).

| Star ID | $T_{\rm eff}$ (K) | $\log g$ | [Fe/H] | $v_{\rm mic}$ (km s$^{-1}$) |
|---|---|---|---|---|
| ra_0103-7050_s163 | $6300 \pm 30$ | $4.49 \pm 0.03$ | $-2.34 \pm 0.06$ | $1.25 \pm 0.01$ |
| ra_0834-5220_s316 | $6000 \pm 300$ | $4.29 \pm 0.07$ | $-2.62 \pm 0.14$ | $1.25 \pm 0.06$ |
| ra_1604-2712_s24 | $5000 \pm 40$ | $1.89 \pm 0.09$ | $-2.53 \pm 0.06$ | $2.01 \pm 0.08$ |
| ra_1604-2712_s292 | $5780 \pm 50$ | $3.41 \pm 0.21$ | $-3.10 \pm 0.07$ | $1.55 \pm 0.11$ |
| ra_1624-2150_s278 | $6400 \pm 200$ | $3.97 \pm 0.20$ | $-2.67 \pm 0.06$ | $1.49 \pm 0.13$ |
| ra_1633-2814_s130 | $5000 \pm 200$ | $1.85 \pm 0.78$ | $-3.35 \pm 0.07$ | $2.04 \pm 0.65$ |
| ra_1633-2814_s284 | $5300 \pm 100$ | $2.45 \pm 0.67$ | $-3.89 \pm 0.07$ | $1.81 \pm 0.48$ |
| ra_1648-0653_s38 | $6000 \pm 200$ | $3.65 \pm 0.16$ | $-2.48 \pm 0.06$ | $1.53 \pm 0.12$ |
| ra_1656-1433_s143 | $4800 \pm 200$ | $1.41 \pm 0.52$ | $-2.86 \pm 0.06$ | $2.21 \pm 0.52$ |
| ra_1658-2454_s22 | $4850 \pm 90$ | $2.17 \pm 0.31$ | $-2.90 \pm 0.08$ | $1.67 \pm 0.22$ |
| ra_1659-2154_s114 | $4960 \pm 70$ | $1.73 \pm 0.16$ | $-3.73 \pm 0.07$ | $2.10 \pm 0.16$ |
| ra_1709-2130_s102 | $5000 \pm 100$ | $1.85 \pm 0.50$ | $-3.00 \pm 0.06$ | $2.03 \pm 0.42$ |
| ra_1752-4300_s214 | $4850 \pm 40$ | $1.55 \pm 0.16$ | $-2.77 \pm 0.05$ | $2.15 \pm 0.15$ |
| ra_1752-4300_s269 | $4920 \pm 30$ | $1.68 \pm 0.56$ | $-2.85 \pm 0.05$ | $2.10 \pm 0.50$ |
| ra_1752-4300_s6 | $4860 \pm 50$ | $1.55 \pm 0.18$ | $-3.01 \pm 0.06$ | $2.16 \pm 0.17$ |
| ra_1853-3255_s45 | $5030 \pm 20$ | $2.38 \pm 0.16$ | $-2.94 \pm 0.07$ | $1.68 \pm 0.11$ |

*Note.* The full table is available in machine-readable format in the electronic version of the paper.

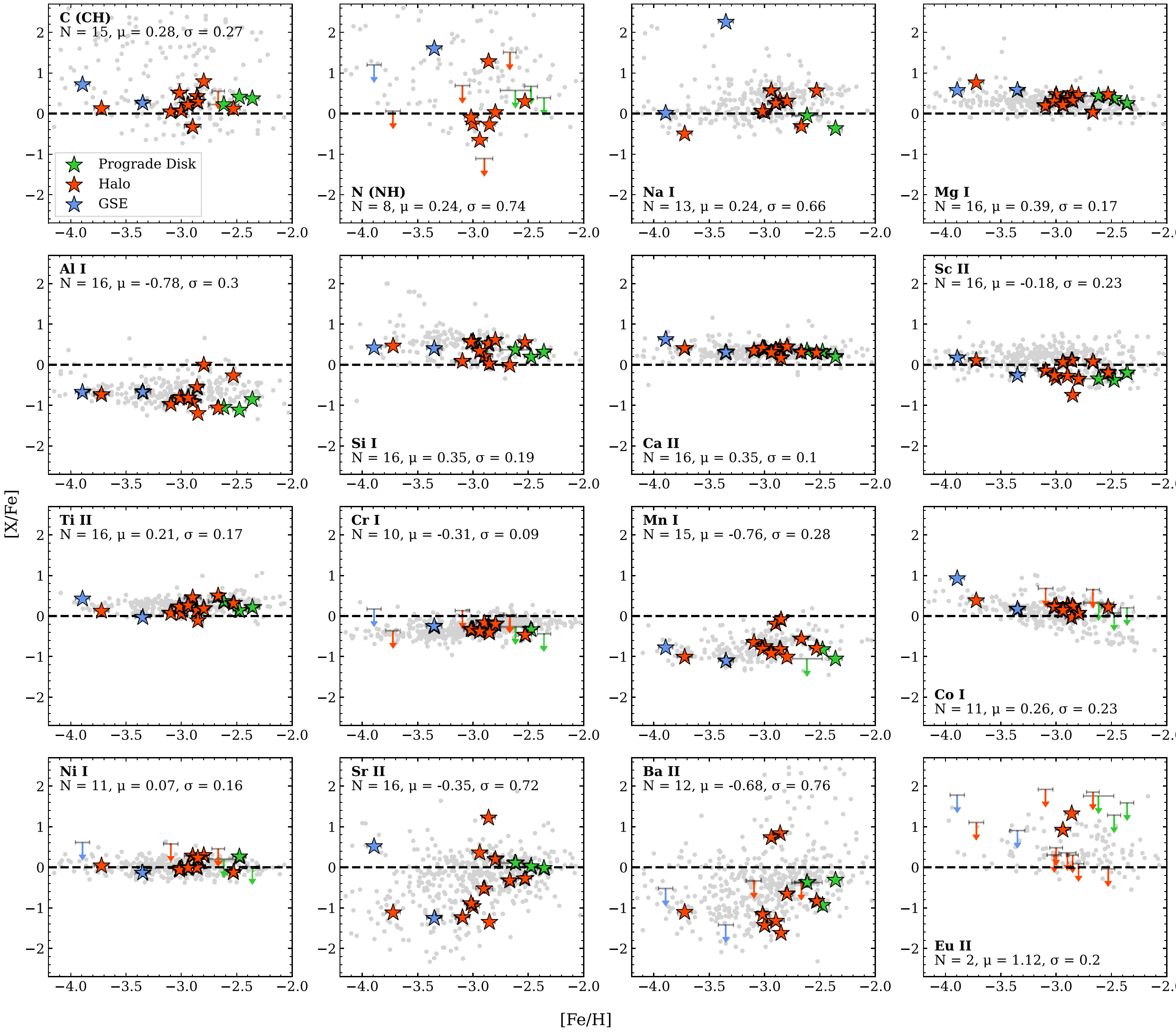

**Figure 2.** Chemical abundances for our sample against high-resolution literature values (grey points) (D. Yong et al. 2013, 2021a; H. R. Jacobson et al. 2015; A. F. Marino et al. 2019). Each panel represents a different element measured. Those with upper-limits are shown by downward-facing arrows. The number of stars with measurements, alongside their mean and standard deviations, are given on each panel. For C, the abundance measurements have been evolutionary-corrected (to be consistent with literature values). For Ca II, we show the NLTE-corrected abundances, while the literature values were measured from Ca I (not NLTE-corrected).

For C, we were able to successfully measure this element from the CH band (∼4300 Å) for 15 out of the 16 stars. The spectral fits for these, and those where only upper limits were determined, are shown in Figs A1 and A2. Evolutionary mixing corrections were applied to our red giant stars (those with $\log g < 3.0$, see Table 4) using the corrections supplied by V. M. Placco et al. (2014) (see Section 3.2.2). The corrections to the observed [C/Fe] values range from +0.01 to +0.46 dex.

Compared with our literature sample, our corrected C abundances show significantly less scatter than the literature sample ($\sigma_{\rm obs} = 0.27$ dex versus $\sigma_{\rm lit} = 0.96$ dex). Similarly, the mean $[{\rm C/Fe}]_{\rm corr}$, at 0.28 dex is considerably lower than the mean for the literature sample at 0.82 dex. These differences are a consequence of the lack of any significant C-enhancement in our small sample. A larger sample size will likely increase the scatter and mean abundances. For those stars with detections, the mean values across our kinematic groups are consistent within the scatter, again likely due to our small sample. Nevertheless, in Fig. 4, noting the definition of C-enhanced stars as those satisfying [C/Fe] $> 0.7$ (W. Aoki et al. 2007), we identify two C-rich stars in our sample. A third has errors on their [C/Fe] values that might place them in the C-rich region. One star is C-depleted ([C/Fe] $< 0$, whilst an additional two might also be in this category, given the uncertainty in their [C/Fe] values). The remaining stars have $0 <$ [C/Fe] $< 0.7$ and are C-normal.

For N, we detected the NH band at ∼3360 Å for eight stars from our sample. The spectral fits are shown in Figs B1 and B2, along with the upper limits on [N/Fe] for the non-detections. Our sample has $\sigma_{\rm obs} = 0.74$ dex, which agrees with literature scatter at $\sigma_{\rm lit} = 0.91$ dex. We have two stars with [N/Fe] $> 1.0$:

**Table 5.** Metallicity and abundance measurements for our sample. Those with no detections have their upper-limits represented instead. Note that stars ra_1604-2712_s292, ra_1648-0653_s38 and ra_1752-4300_s269 have no [Na/Fe] measurements due to contamination from interstellar Na absorption.

| star | [Fe/H]<br>[Sc/Fe] | $[C/Fe]_{raw}$<br>[Ti/Fe] | $[C/Fe]_{corr}$<br>[Cr/Fe] | [N/Fe]<br>[Mn/Fe] | [Na/Fe]<br>[Co/Fe] | [Mg/Fe]<br>[Ni/Fe] | [Al/Fe]<br>[Sr/Fe] | [Si/Fe]<br>[Ba/Fe] | [Ca/Fe]<br>[Eu/Fe] |
|---|---|---|---|---|---|---|---|---|---|
| ra_0103-7050_s163 | $-2.36 \pm 0.06$ | $0.37 \pm 0.08$ | $0.37 \pm 0.08$ | $< 0.39$ | $-0.37 \pm 0.06$ | $0.25 \pm 0.05$ | $-0.85 \pm 0.03$ | $0.31 \pm 0.06$ | $0.20 \pm 0.03$ |
| | $-0.20 \pm 0.04$ | $0.21 \pm 0.05$ | $< -0.44$ | $-1.06 \pm 0.04$ | $< 0.20$ | $< 0.01$ | $-0.02 \pm 0.07$ | $-0.32 \pm 0.06$ | $< 1.59$ |
| ra_0834-5220_s316 | $-2.62 \pm 0.14$ | $0.22 \pm 0.07$ | $0.22 \pm 0.07$ | $< 0.57$ | $-0.05 \pm 0.04$ | $0.43 \pm 0.06$ | $-1.05 \pm 0.03$ | $0.37 \pm 0.06$ | $0.35 \pm 0.03$ |
| | $-0.34 \pm 0.04$ | $0.36 \pm 0.03$ | $< -0.27$ | $< -1.05$ | $< 0.32$ | $< 0.20$ | $0.10 \pm 0.07$ | $-0.37 \pm 0.12$ | $< 1.76$ |
| ra_1604-2712_s24 | $-2.53 \pm 0.06$ | $-0.00 \pm 0.08$ | $0.12 \pm 0.08$ | $0.30 \pm 0.09$ | $0.56 \pm 0.07$ | $0.46 \pm 0.07$ | $-0.27 \pm 0.06$ | $0.55 \pm 0.07$ | $0.29 \pm 0.02$ |
| | $-0.19 \pm 0.09$ | $0.31 \pm 0.04$ | $-0.47 \pm 0.05$ | $-0.79 \pm 0.10$ | $0.22 \pm 0.06$ | $-0.13 \pm 0.04$ | $-0.28 \pm 0.09$ | $-0.84 \pm 0.04$ | $< -0.04$ |
| ra_1604-2712_s292 | $-3.10 \pm 0.07$ | $0.04 \pm 0.10$ | $0.04 \pm 0.10$ | $< 0.69$ | $< 0.68$ | $0.19 \pm 0.05$ | $-0.97 \pm 0.03$ | $0.09 \pm 0.04$ | $0.34 \pm 0.02$ |
| | $-0.15 \pm 0.04$ | $0.07 \pm 0.04$ | $< 0.13$ | $-0.65 \pm 0.04$ | | $< 0.58$ | $-1.24 \pm 0.04$ | $< -0.33$ | $< 1.92$ |
| ra_1624-2150_s278 | $-2.67 \pm 0.06$ | $< 0.55$ | $< 0.55$ | $< 1.51$ | $-0.31 \pm 0.02$ | $0.03 \pm 0.04$ | $-1.06 \pm 0.03$ | $-0.01 \pm 0.03$ | $0.31 \pm 0.01$ |
| | $0.08 \pm 0.04$ | $0.50 \pm 0.04$ | $< 0.02$ | $-0.56 \pm 0.04$ | $< 0.65$ | $< 0.46$ | $-0.33 \pm 0.05$ | $< -0.38$ | $< 1.85$ |
| ra_1633-2814_s130 | $-3.35 \pm 0.07$ | $0.23 \pm 0.08$ | $0.26 \pm 0.08$ | $1.60 \pm 0.10$ | $2.26 \pm 0.07$ | $0.58 \pm 0.04$ | $-0.67 \pm 0.04$ | $0.40 \pm 0.08$ | $0.31 \pm 0.03$ |
| | $-0.26 \pm 0.04$ | $-0.03 \pm 0.04$ | $-0.25 \pm 0.04$ | $-1.10 \pm 0.05$ | $0.17 \pm 0.04$ | $-0.14 \pm 0.03$ | $-1.25 \pm 0.05$ | $< -1.42$ | $< 0.90$ |
| ra_1633-2814_s284 | $-3.89 \pm 0.06$ | $0.71 \pm 0.08$ | $0.71 \pm 0.08$ | $< 1.20$ | $0.01 \pm 0.11$ | $0.57 \pm 0.04$ | $-0.67 \pm 0.03$ | $0.42 \pm 0.06$ | $0.63 \pm 0.02$ |
| | $0.17 \pm 0.04$ | $0.42 \pm 0.04$ | $< 0.17$ | $-0.78 \pm 0.05$ | $0.92 \pm 0.04$ | $< 0.61$ | $0.51 \pm 0.09$ | $< -0.52$ | $< 1.78$ |
| ra_1648-0653_s38 | $-2.48 \pm 0.06$ | $0.41 \pm 0.08$ | $0.41 \pm 0.08$ | $< 0.67$ | $< 0.08$ | $0.37 \pm 0.05$ | $-1.12 \pm 0.03$ | $0.19 \pm 0.06$ | $0.32 \pm 0.02$ |
| | $-0.38 \pm 0.04$ | $0.14 \pm 0.04$ | $-0.33 \pm 0.03$ | $-0.82 \pm 0.04$ | | $0.26 \pm 0.03$ | $0.02 \pm 0.08$ | $-0.94 \pm 0.12$ | $< 1.28$ |
| ra_1656-1433_s143 | $-2.86 \pm 0.06$ | $-0.04 \pm 0.22$ | $0.42 \pm 0.22$ | $1.28 \pm 0.10$ | $0.32 \pm 0.07$ | $0.50 \pm 0.05$ | $-0.56 \pm 0.06$ | $0.52 \pm 0.09$ | $0.42 \pm 0.02$ |
| | $0.11 \pm 0.09$ | $0.14 \pm 0.04$ | $-0.23 \pm 0.05$ | $-0.82 \pm 0.08$ | $-0.03 \pm 0.05$ | $-0.01 \pm 0.03$ | $1.22 \pm 0.03$ | $0.83 \pm 0.08$ | $1.32 \pm 0.03$ |
| ra_1658-2454_s22 | $-2.90 \pm 0.08$ | $-0.33 \pm 0.08$ | $-0.33 \pm 0.08$ | $< -1.11$ | $0.25 \pm 0.07$ | $0.37 \pm 0.06$ | $-0.91 \pm 0.05$ | $0.21 \pm 0.09$ | $0.36 \pm 0.03$ |
| | $-0.28 \pm 0.08$ | $0.45 \pm 0.04$ | $-0.17 \pm 0.05$ | $-0.19 \pm 0.11$ | $0.26 \pm 0.06$ | $0.28 \pm 0.05$ | $-0.53 \pm 0.10$ | $-1.32 \pm 0.03$ | $< 0.36$ |
| ra_1659-2154_s114 | $-3.72 \pm 0.07$ | $0.00 \pm 0.09$ | $0.12 \pm 0.09$ | $< 0.06$ | $-0.50 \pm 0.03$ | $0.76 \pm 0.06$ | $-0.73 \pm 0.03$ | $0.46 \pm 0.06$ | $0.40 \pm 0.03$ |
| | $0.11 \pm 0.04$ | $0.12 \pm 0.03$ | $< -0.36$ | $-1.01 \pm 0.05$ | $0.38 \pm 0.04$ | $0.04 \pm 0.03$ | $-1.12 \pm 0.04$ | $-1.11 \pm 0.07$ | $< 1.11$ |
| ra_1709-2130_s102 | $-3.00 \pm 0.06$ | $0.01 \pm 0.08$ | $0.07 \pm 0.08$ | $-0.25 \pm 0.09$ | $0.04 \pm 0.05$ | $0.47 \pm 0.05$ | $-0.81 \pm 0.05$ | $0.58 \pm 0.08$ | $0.39 \pm 0.02$ |
| | $-0.33 \pm 0.06$ | $0.06 \pm 0.05$ | $-0.34 \pm 0.04$ | $-0.75 \pm 0.06$ | $0.25 \pm 0.05$ | $-0.02 \pm 0.03$ | $-0.97 \pm 0.08$ | $-1.43 \pm 0.07$ | $< 0.48$ |
| ra_1752-4300_s214 | $-2.80 \pm 0.05$ | $0.51 \pm 0.08$ | $0.79 \pm 0.08$ | $0.03 \pm 0.10$ | $0.31 \pm 0.07$ | $0.45 \pm 0.06$ | $-0.00 \pm 0.07$ | $0.61 \pm 0.05$ | $0.45 \pm 0.02$ |
| | $-0.35 \pm 0.13$ | $0.18 \pm 0.04$ | $-0.20 \pm 0.05$ | $-1.01 \pm 0.11$ | $0.07 \pm 0.05$ | $0.29 \pm 0.04$ | $0.20 \pm 0.08$ | $-0.65 \pm 0.04$ | $< 0.09$ |
| ra_1752-4300_s269 | $-2.85 \pm 0.05$ | $0.06 \pm 0.09$ | $0.28 \pm 0.09$ | $-0.27 \pm 0.09$ | $0.25 \pm 0.05$ | $0.32 \pm 0.05$ | $-1.20 \pm 0.04$ | $0.03 \pm 0.12$ | $0.17 \pm 0.02$ |
| | $-0.75 \pm 0.07$ | $-0.11 \pm 0.05$ | $-0.41 \pm 0.05$ | $-0.09 \pm 0.13$ | | $0.23 \pm 0.04$ | $-1.35 \pm 0.07$ | $-1.62 \pm 0.04$ | $< 0.30$ |
| ra_1752-4300_s6 | $-3.02 \pm 0.06$ | $0.19 \pm 0.23$ | $0.50 \pm 0.23$ | $-0.10 \pm 0.10$ | $0.06 \pm 0.05$ | $0.30 \pm 0.06$ | $-0.82 \pm 0.05$ | $0.56 \pm 0.10$ | $0.41 \pm 0.02$ |
| | $-0.26 \pm 0.09$ | $0.23 \pm 0.04$ | $-0.33 \pm 0.05$ | $-0.81 \pm 0.08$ | $0.23 \pm 0.05$ | $-0.06 \pm 0.03$ | $-0.89 \pm 0.08$ | $-1.16 \pm 0.04$ | $< 0.31$ |
| ra_1853-3255_s45 | $-2.94 \pm 0.07$ | $0.20 \pm 0.08$ | $0.21 \pm 0.08$ | $-0.66 \pm 0.10$ | $0.57 \pm 0.07$ | $0.21 \pm 0.07$ | $-0.82 \pm 0.04$ | $0.34 \pm 0.10$ | $0.30 \pm 0.03$ |
| | $0.07 \pm 0.09$ | $0.26 \pm 0.05$ | $-0.38 \pm 0.04$ | $-0.91 \pm 0.07$ | $0.14 \pm 0.05$ | $-0.01 \pm 0.03$ | $0.36 \pm 0.07$ | $0.73 \pm 0.02$ | $0.92 \pm 0.04$ |

*Note.* The full table is available in machine-readable format in the electronic version of the paper.

halo star ra_1656-1433_s143 with [N/Fe] $= 1.28 \pm 0.10$, and GSE star ra_1633-2814_s130 with [N/Fe] $= 1.60 \pm 0.10$. High N can come from evolutionary CNO cycle mixing, which is the likely explanation for ra_1656-1433_s143 (having a high evolutionary correction to the C abundance of 0.46 dex), but that's unlikely to be the case for ra_1633-2814_s130 given it is negligible evolutionary mixing correction for [C/Fe] (0.03 dex). Therefore, ra_1656-1433_s143 is likely a NEMP star as per definition by J. A. Johnson et al. (2007) ([N/Fe] $> 0.5$ and [C/N] $< -0.5$). We show this in Fig. 5 for our sample, alongside the values in Table 6. See also Section 5.1 for discussion on GSE star ra_1633-2814_s130. Alongside the detections, we have eight non-detections, with halo star ra_1658-2454_s22 having a low upper-limit of [N/Fe] $< -1.11$. No star in the comparison sample has an N value/upper-limit this low. We will discuss this star more in Section 5.3.

#### 4.2.2 *Odd-Z elements*

For the odd-Z[6] elements, we measured Na, Al, and Sc in this study. The production sites for these elements are through hydrostatic burning in massive stars, together with explosive nucleosynthesis.

We successfully measured Na in 13 out of the 16 stars from the Na D-lines at 5889.95 and 5895.92 Å. Interstellar Na absorption features from intervening line-of-sight gas clouds contaminated the stellar features for the remaining three stars. The scatter of our sample is $\sigma_{obs} = 0.66$ dex, broadly consistent with that measured in the comparison literature at $\sigma_{lit} = 0.46$ dex. Our 1D LTE [Na/Fe] abundances derived from the Na D lines are subject to corrections derived from more realistic NLTE models (e.g. K. Lind et al. 2011, 2022; S. A. Alexeeva, Y. V. Pakhomov & L. I. Mashonkina 2014; L. Mashonkina et al. 2023). Using the 5889.95 Å Na absorption feature, a typical star (with $T_{eff} = 5000$ K, $\log g = 2.0$, $v_{mic} = 2.0\,km\,s^{-1}$, and [Fe/H] $= -3.0$) has the following corrections from the INSPECT database[7]: $-0.14$ dex for [Na/Fe] $= -0.6$, $-0.36$ dex for [Na/Fe] $= 0.0$, and $-0.57$ dex for [Na/Fe] $= 0.6$.

Of particular interest is the star ra_1633-2814_s130, which has [Na/Fe] $= 2.26 \pm 0.07$, placing it amongst the most Na-rich stars known among metal-poor stars (see panel in Fig. 2). Compared to our sample's mean, $0.07 \pm 0.35$ dex, ra_1633-2814_s130 is significantly higher, higher also than the comparison literature's mean at $0.32 \pm 0.45$ dex. This star, alongside being the NEMP identified in the previous section, also has a strong Li enhancement, which will be discussed further in Section 5.1.

For Al, all 16 stars had detections from the 3961.52 Å line, with a scatter of $\sigma_{obs} = 0.30$ dex. This is consistent with the scatter of the literature at $\sigma_{lit} = 0.30$ dex. Within errors, the mean values across the prograde disc ($-1.01 \pm 0.14$ dex), halo ($-0.74 \pm 0.35$ dex), and GSE ($-0.671 \pm 0.006$ dex) agree, with a possible small offset present between the prograde disc and GSE populations.

[6]Z being atomic number.

[7]https://www.inspect-stars.com/

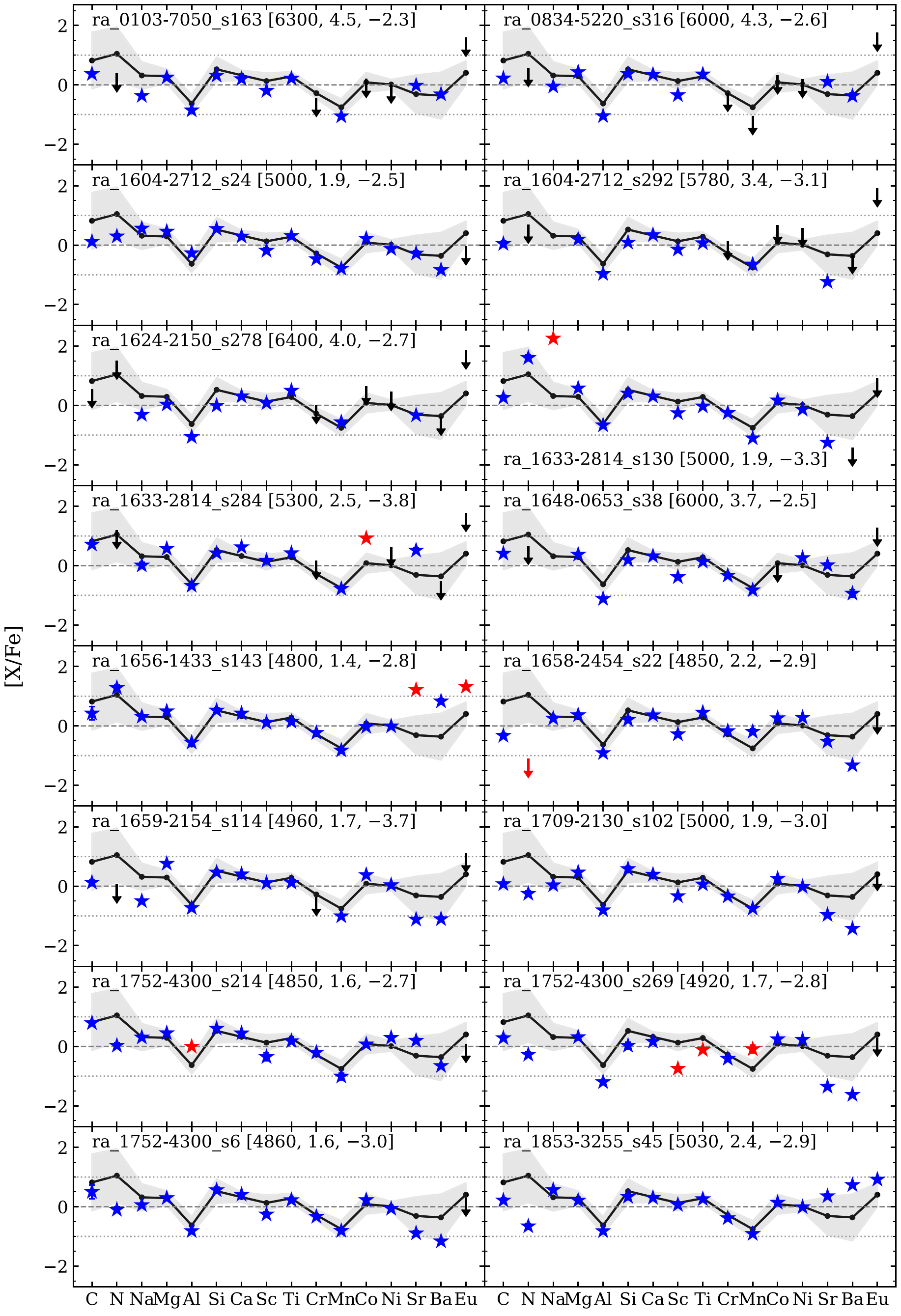

**Figure 3.** Abundance patterns [X/Fe] for each measured element for our 16 metal-poor stars. In each panel, the black dots correspond to the mean literature values (D. Yong et al. 2013; H. R. Jacobson et al. 2015; A. F. Marino et al. 2019; D. Yong et al. 2021a), with the star symbols corresponding to the abundance value for that given star. Those coloured blue indicate measurements within $2\sigma$ of the mean value, otherwise they are red. Downward-facing arrows are upper-limits, and if these are $2\sigma$ away from the mean, they are also shown in red. Otherwise they are black. The shaded region is the standard deviation of the literature values for the particular element. Horizontal grey dashed lines are given at [X/Fe] = 0 and [X/Fe] = ±1.0. The star's $T_{\mathrm{eff}}$, log *g*, and [Fe/H]are given in each panel.

For Sc, we successfully measured an abundance from the 4246.82 Å line for all 16 stars. We note that this line is located near a CH molecular band head, but the line is largely unblended even at our resolution. For our sample, we find a scatter of $\sigma_{\mathrm{obs}} = 0.23$ dex, in excellent agreement with the literature at $\sigma_{\mathrm{lit}} = 0.27$ dex. The mean values between the prograde disc ($-0.31 \pm 0.10$ dex), halo ($-0.18 \pm 0.26$), and the GSE ($-0.04 \pm 0.30$) are consistent within errors. Interestingly, the prograde disc has the smallest scatter amongst the three kinematic regions.

### 4.2.3 $\alpha$-elements

The $\alpha$-elements we measured in this study were Mg, Si, Ca, and Ti. These elements are made primarily through stellar nucleosynthesis and ejected through core-collapse supernovae (Type II). All of our stars had detections for each of the $\alpha$-elements. The average $\alpha$-element abundance was calculated from the weighted mean between Mg, Si, Ca, and Ti. As shown in Fig. 6 and in Table 7, we have two stars with $[\alpha/\mathrm{Fe}] > 0.4$ (and one within

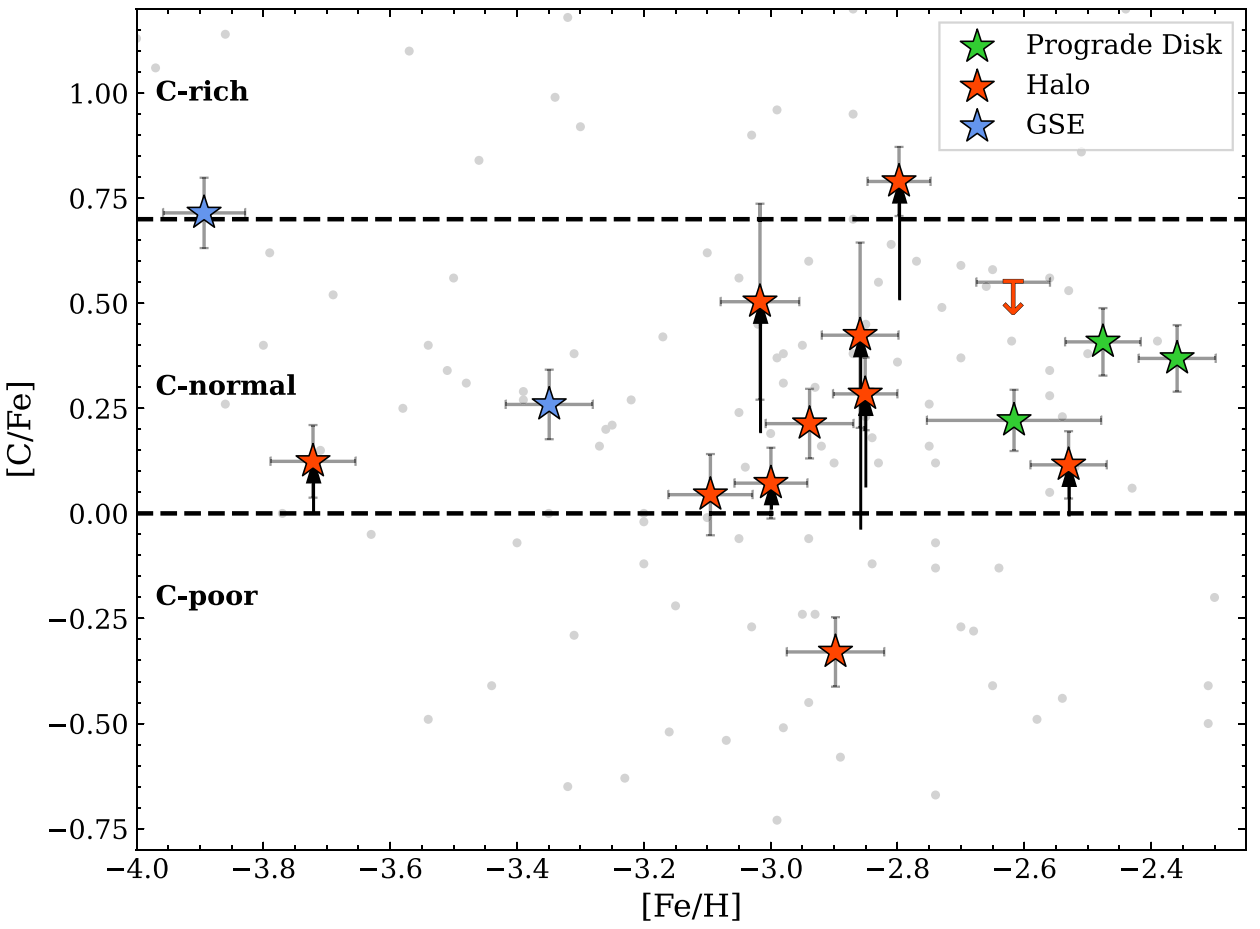

**Figure 4.** Evolutionary-corrected C abundances for our sample, with literature shown by the light grey points. Plot is separated into three regions: C-rich ([C/Fe] > 0.7), C-normal (0 < [C/Fe] ≤ 0.7), and C-poor ([C/Fe] ≤ 0). For the stars with detections: two stars are C-rich (one within errors), one is C-poor (one within errors), with the rest C-normal. For stars with evolutionary corrections, their raw measured value is shown by the black arrows.

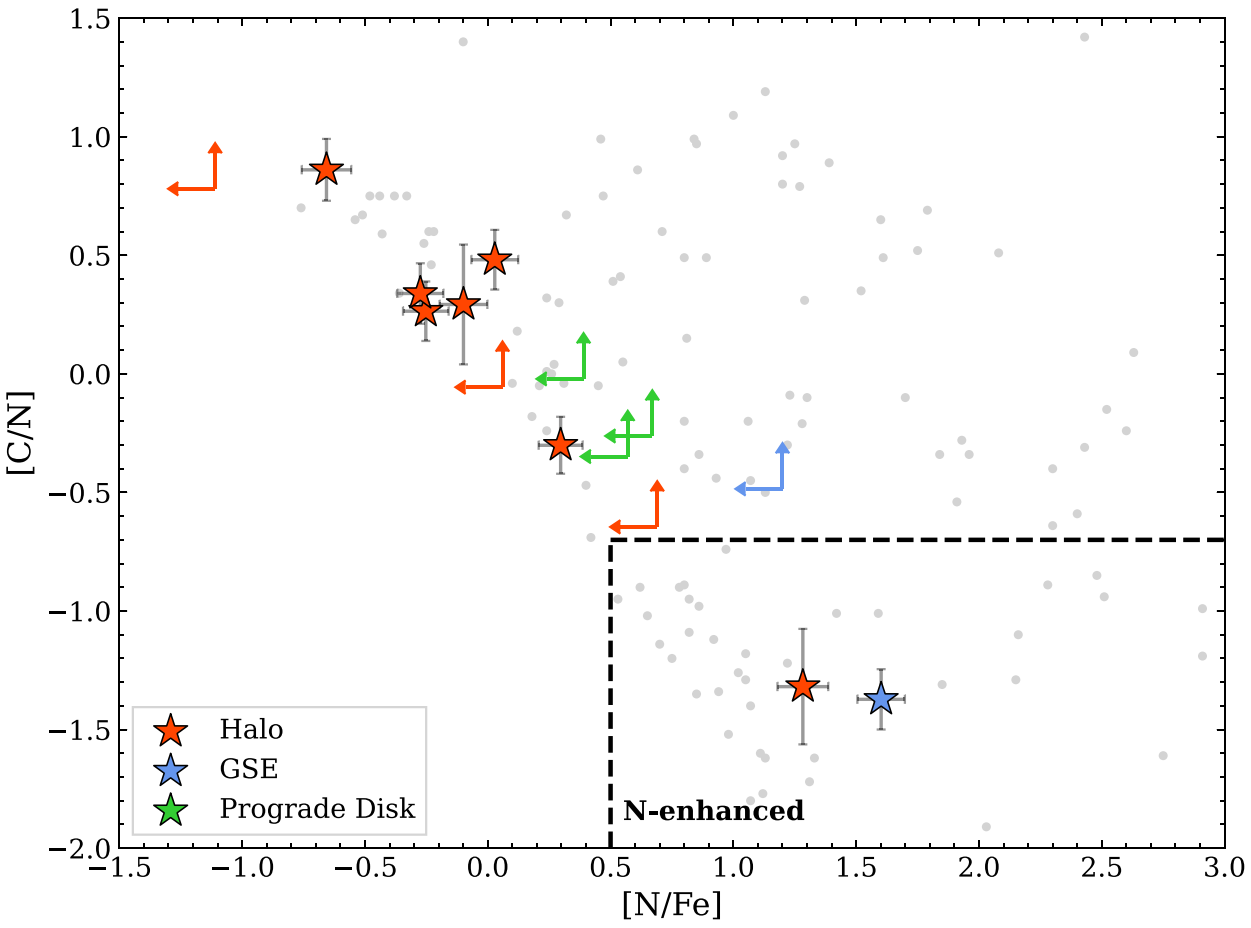

**Figure 5.** Identifying stars as NEMPs by comparing [N/Fe] with [C/N] (see Table 6 for their values). For consistency, we use the uncorrected C abundances when determining [C/N]. Literature is plotted underneath in light grey. Those with upper limits in [N/Fe] (represented by leftward-facing arrows), but detections in [C/Fe] have lower limits shown for [C/N] (represented by upward-facing arrows). Stars with upper limits in both [N/Fe] and [C/Fe] are not included. Stars are assigned as NEMP if [N/Fe] > 0.5 and [C/N] < −0.5 (J. A. Johnson et al. 2007), as shown by the dashed rectangle on the plot. From this, we identify two possible NEMP stars: one likely due to it having high-evolutionary mixing corrections for [C/Fe], raising C and lowering N (0.46 dex), with the other unlikely given it has low evolutionary corrections (0.03 dex).

error): GSE star ra_1633-2814_s284 with [$\alpha$/Fe] = 0.56 ± 0.02, and halo star ra_1752-4300_s214 with [$\alpha$/Fe] = 0.42 ± 0.02. We have no stars with [$\alpha$/Fe] < 0.1. Given the small sample size, the mean values with the prograde disc (0.29 ± 0.07), the halo (0.32 ± 0.08), and the GSE (0.44 ± 0.17) are consistent with each other.

Looking at the individual $\alpha$-elements, we measured Mg from the Mg-triplet across 5167–5183 Å, and the line at 8806.76 Å. The

**Table 6.** [N/Fe] and [C/N] values for our sample. Note that we use the uncorrected C values when calculating [C/N]. Those with upper limits in [N/Fe], but detections in [C/Fe] will have a corresponding lower limit for [C/N]. Those with upper limits in both [N/Fe] and [C/Fe] are not included. Star ra_1656-1433_s143 is likely a NEMP, whilst star ra_1658-2454_s22 has exceptionally low [N/Fe] upper limits. This latter star will be discussed further in Section 5.3.

| Star | [N/Fe] | [C/N] |
|---|---|---|
| ra_0103-7050_s163 | <0.39 | > −0.02 |
| ra_0834-5220_s316 | <0.57 | > −0.35 |
| ra_1604-2712_s24 | 0.30 ± 0.09 | −0.30 ± 0.12 |
| ra_1604-2712_s292 | <0.69 | > −0.65 |
| ra_1633-2814_s130 | 1.60 ± 0.10 | −1.37 ± 0.13 |
| ra_1633-2814_s284 | <1.20 | > −0.49 |
| ra_1648-0653_s38 | <0.67 | > −0.26 |
| ra_1656-1433_s143 | 1.28 ± 0.10 | −1.32 ± 0.24 |
| ra_1658-2454_s22 | < −1.11 | >0.78 |
| ra_1659-2154_s114 | <0.06 | > −0.06 |
| ra_1709-2130_s102 | −0.25 ± 0.09 | 0.26 ± 0.13 |
| ra_1752-4300_s214 | 0.03 ± 0.10 | 0.48 ± 0.13 |
| ra_1752-4300_s269 | −0.27 ± 0.09 | 0.34 ± 0.13 |
| ra_1752-4300_s6 | −0.10 ± 0.10 | 0.29 ± 0.25 |
| ra_1853-3255_s45 | −0.66 ± 0.10 | 0.86 ± 0.13 |

*Note.* The full table is available in machine-readable format in the electronic version of the paper.

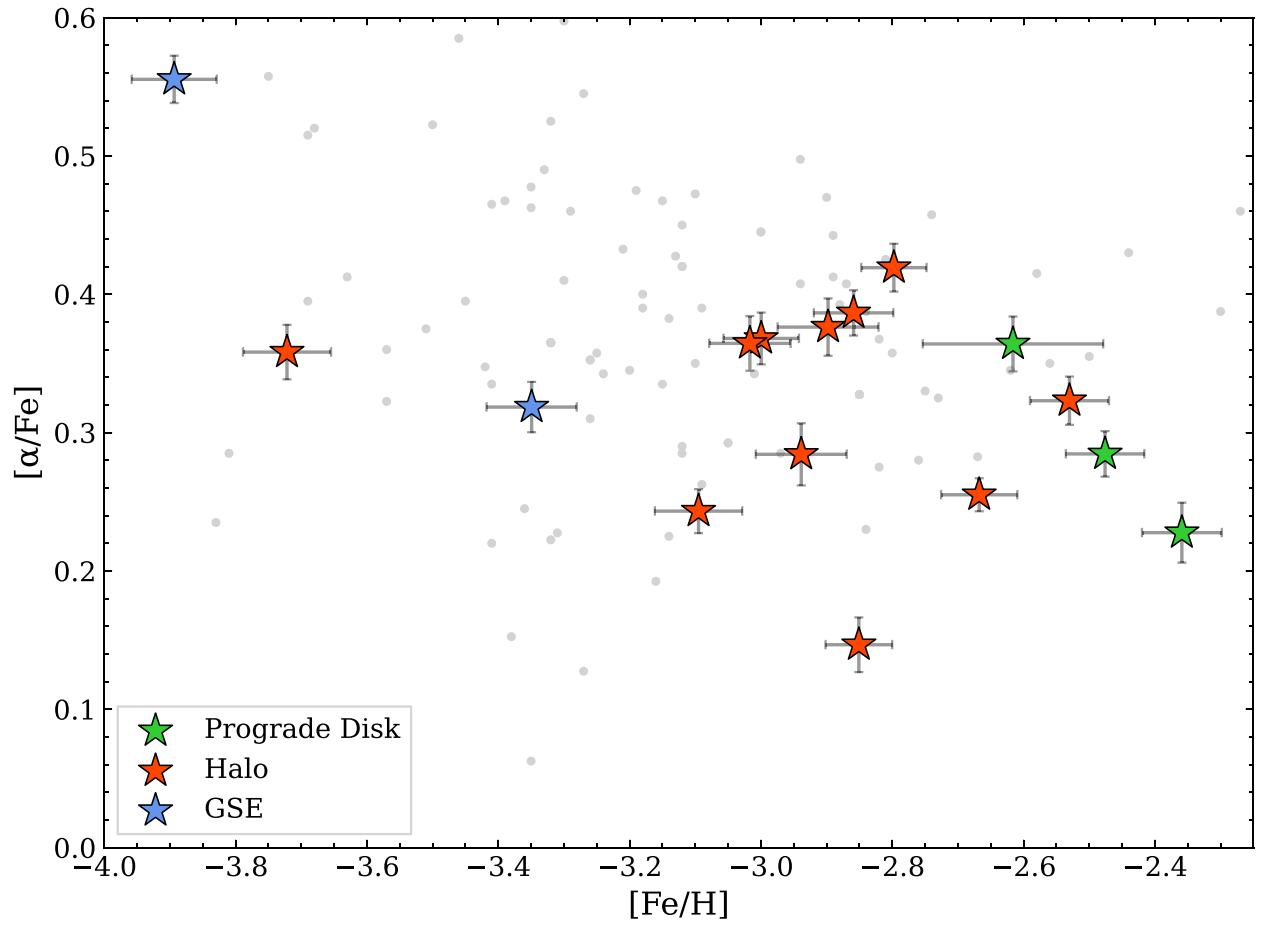

**Figure 6.** $\alpha$-abundances for our sample (see Table 7 for their values). Our $\alpha$-elements include Ca, Mg, Si, and Ti. For consistency, the literature $\alpha$-abundances (shown in light grey) also comprised of the same elements we used.

scatter of our data at $\sigma_{obs}$ = 0.17 dex is in strong agreement with the comparison literature at $\sigma_{lit}$ = 0.24 dex. There is little variation amongst the different kinematic groups.

Si was measured from a single line at 3905.52 Å. We see a small spread at $\sigma_{obs}$ = 0.19 dex than with the comparison literature at $\sigma_{lit}$ = 0.41 dex. This is due to the larger spread seen for stars at [Fe/H] < −3.5. The halo sample has the largest scatter at $\sigma_{halo}$ = 0.24 dex compared with the prograde disc ($\sigma_{pro}$ = 0.091 dex) and GSE ($\sigma_{gse}$ = 0.013 dex) samples, though this is likely due to low-number statistics.

Ca for our sample was measured from the Ca II triplet located at wavelengths 8498.23, 8542.31, and 8662.36 Å. We encountered issues trying to fit the Ca I 4300.313 Å line reliably due to poor SNR in the region, so we only report the findings from Ca II (which has been NLTE corrected and is shown in the Ca II panel

**Table 7.** [$\alpha$/Fe] values for our sample, derived from the weighted mean of Ca, Mg, Si, and Ti abundances.

| Star | [Fe/H] | [$\alpha$/Fe] |
|---|---|---|
| ra_0103-7050_s163 | $-2.36 \pm 0.06$ | $0.23 \pm 0.02$ |
| ra_0834-5220_s316 | $-2.62 \pm 0.14$ | $0.36 \pm 0.02$ |
| ra_1604-2712_s24 | $-2.53 \pm 0.06$ | $0.32 \pm 0.02$ |
| ra_1604-2712_s292 | $-3.10 \pm 0.07$ | $0.24 \pm 0.02$ |
| ra_1624-2150_s278 | $-2.67 \pm 0.06$ | $0.26 \pm 0.01$ |
| ra_1633-2814_s130 | $-3.35 \pm 0.07$ | $0.32 \pm 0.02$ |
| ra_1633-2814_s284 | $-3.89 \pm 0.06$ | $0.56 \pm 0.02$ |
| ra_1648-0653_s38 | $-2.48 \pm 0.06$ | $0.28 \pm 0.02$ |
| ra_1656-1433_s143 | $-2.86 \pm 0.06$ | $0.39 \pm 0.02$ |
| ra_1658-2454_s22 | $-2.90 \pm 0.08$ | $0.38 \pm 0.02$ |
| ra_1659-2154_s114 | $-3.72 \pm 0.07$ | $0.36 \pm 0.02$ |
| ra_1709-2130_s102 | $-3.00 \pm 0.06$ | $0.37 \pm 0.02$ |
| ra_1752-4300_s214 | $-2.80 \pm 0.05$ | $0.42 \pm 0.02$ |
| ra_1752-4300_s269 | $-2.85 \pm 0.05$ | $0.15 \pm 0.02$ |
| ra_1752-4300_s6 | $-3.02 \pm 0.06$ | $0.36 \pm 0.02$ |
| ra_1853-3255_s45 | $-2.94 \pm 0.07$ | $0.28 \pm 0.02$ |

*Note.* The full table is available in machine-readable format in the electronic version of the paper.

in Fig. 2). We note that the literature Ca values come from Ca I, so we are comparing our Ca II with their Ca I values. Our data has low scatter of $\sigma_{\rm obs} = 0.10$ dex, consistent with the comparison literature at $\sigma_{\rm lit} = 0.17$ dex. Our [Ca/Fe] measurements show a slight increase in [Ca/Fe] with decreasing [Fe/H], consistent with the trend present in the literature values. Amongst the kinematic groups, the GSE has the highest mean abundance ($0.47 \pm 0.17$ dex) compared to the halo ($0.35 \pm 0.08$ dex) and the prograde disc ($0.29 \pm 0.09$ dex), but the values are consistent with each other, given the uncertainties.

Ti was measured using the lines at 3913.46, 4468.49, and 4501.27 Å, giving us a scatter of $\sigma_{\rm obs} = 0.17$ dex. This is in excellent agreement with the comparison literature, having a scatter $\sigma_{\rm lit} = 0.16$ dex. The mean abundances of the halo ($0.20 \pm 0.17$ dex), the GSE ($0.20 \pm 0.32$ dex), and the prograde disc ($0.24 \pm 0.11$ dex) samples are consistent with each other.

### 4.2.4 *Iron peak elements*

The iron peak elements studied in this work included Cr, Mn, Co, and Ni. At low metallicities, these are primarily produced by Type II supernova, while at later times and higher metallicities, Type Ia supernovae dominate their production.

We successfully measured Cr in 10 of our stars from the three Cr I lines across the wavelength region 5200–5210 Å. We have minimal scatter for our sample at $\sigma_{\rm obs} = 0.09$ dex, lower than the comparison literature at $\sigma_{\rm lit} = 0.18$ dex. There is a slight decreasing trend with decreasing metallicity seen in the literature which is difficult to see in our data (due to the lack of detections in stars with [Fe/H] $< -3.0$). NLTE calculations for Cr I indicate that this trend is not physical (e.g. M. Bergemann & G. Cescutti 2010). With the lack of stars with detections in both the GSE and prograde disc samples, we are unable to comment on any kinematic-based trends.

For Mn, we successfully measured an abundance for 15 of our stars using the three Mn I lines at 4026–4038 Å. We have a scatter of $\sigma_{\rm obs} = 0.28$ dex, again in excellent agreement with the comparison literature at $\sigma_{\rm lit} = 0.29$ dex. The mean abundances across the prograde disc ($-0.94 \pm 0.16$ dex), the halo ($-0.69 \pm 0.31$ dex), and the GSE ($-0.94 \pm 0.23$) are all consistent within errors amongst each other.

We were able to successfully measure Co in 11 of our 16 stars with the two Co I lines across the wavelength region 4115–4125 Å. We have a scatter of $\sigma_{\rm obs} = 0.23$ dex, in agreement with the comparison literature at $\sigma_{\rm lit} = 0.19$ dex. No prograde disc star has detections for Co, though the mean abundance for the halo ($0.20 \pm 0.12$ dex) is lower and has less scatter than the GSE ($0.55 \pm 0.53$ dex). This is likely due to low number statistics.

We successfully measured Ni in 11 stars from our sample using the line located at 5476.90 Å. We see a tight trend in both our sample and comparison literature, with the scatters $\sigma_{\rm obs} = 0.16$ dex and $\sigma_{\rm lit} = 0.17$ dex in excellent agreement with each other. Given the lack of detections within our three kinematic groups, we cannot comment on the mean abundances and scatter.

### 4.2.5 *Neutron capture elements*

We measured the neutron-capture elements Sr, Ba, and Eu. At solar metallicity, Sr is primarily produced by the slow neutron capture process (s-process), Eu is primarily produced by the rapid neutron capture process (r-process), and Ba is produced by both (J. Simmerer et al. 2004). Their relative abundances therefore allow distinguishing the contributions from the two neutron-capture processes. Further details of the neutron capture processes in their various forms are discussed below.

For Sr, we successfully measured an abundance for all of our stars from the Sr II line at 4077.71 Å. We see that the scatter for Sr changes based on metallicity. For [Fe/H] $< -2.8$, we see significant scatter, where we have $\sigma_{\rm obs,[Fe/H]<-2.8} = 0.90$ dex for our sample, and $\sigma_{\rm lit,[Fe/H]<-2.8} = 0.70$ dex for the comparison literature. For [Fe/H] $\geq -2.8$, this scatter decreases for both at $\sigma_{\rm obs,[Fe/H]\geq-2.8} = 0.21$ dex and $\sigma_{\rm lit,[Fe/H]\geq-2.8} = 0.45$ dex. The reason for this (and how Sr is formed) has not been well understood, but studies like G. Cescutti & C. Chiappini (2014) and T. M. Sitnova et al. (2025) have suggested that at low [Fe/H] it can be formed through the 'early' s-process, a variation of the standard s-process occurring within massive rotating metal-poor stars. Other theories include neutrino-driven winds from a young neutron star (S. E. Woosley & R. D. Hoffman 1992; Y. Z. Qian & G. J. Wasserburg 2007), weak r-process in Type II supernovae (N. Izutani, H. Umeda & N. Tominaga 2009; A. Arcones & J. Bliss 2014), the $\nu p$-process (M. Eichler et al. 2018; S. Ghosh, N. Wolfe & C. Fröhlich 2022), intermediate n-capture process for asymptotic giant branch (AGB) stars with $M < 4\,{\rm M}_\odot$ (A. Choplin, L. Siess & S. Goriely 2021; A. Choplin et al. 2024), and the i-process within massive very metal-poor stars (P. Banerjee, Y.-Z. Qian & A. Heger 2018). The $\sim$2 dex range in [Sr/Fe] values from $-1$ to $+1$ indicates a variety of stochastic enrichment processes. We also have one star appearing to be enhanced in Sr: star ra_1656-1433_s143 at [Sr/Fe] $= 1.22 \pm 0.03$, which will be discussed more in Section 5.2.

The pattern seen in Sr is also reflected in Ba for our 12 detections from the Ba II lines 4554.03, 4934.08, and 6496.90 Å. For [Fe/H] $< -2.8$, we again see significant scatter for both our sample and the comparison literature, where for the observed we find $\sigma_{\rm obs,[Fe/H]<-2.8} = 1.05$ dex, and for the comparison literature we find $\sigma_{\rm lit,[Fe/H]<-2.8} = 0.67$ dex. For [Fe/H] $\geq -2.8$, we determine $\sigma_{\rm obs,[Fe/H]\geq-2.8} = 0.27$ dex and $\sigma_{\rm lit,[Fe/H]\geq-2.8} = 0.85$ dex for the observed and comparison literature data sets, respectively. Like Sr, the process forming Ba at low metallicities is also not understood,

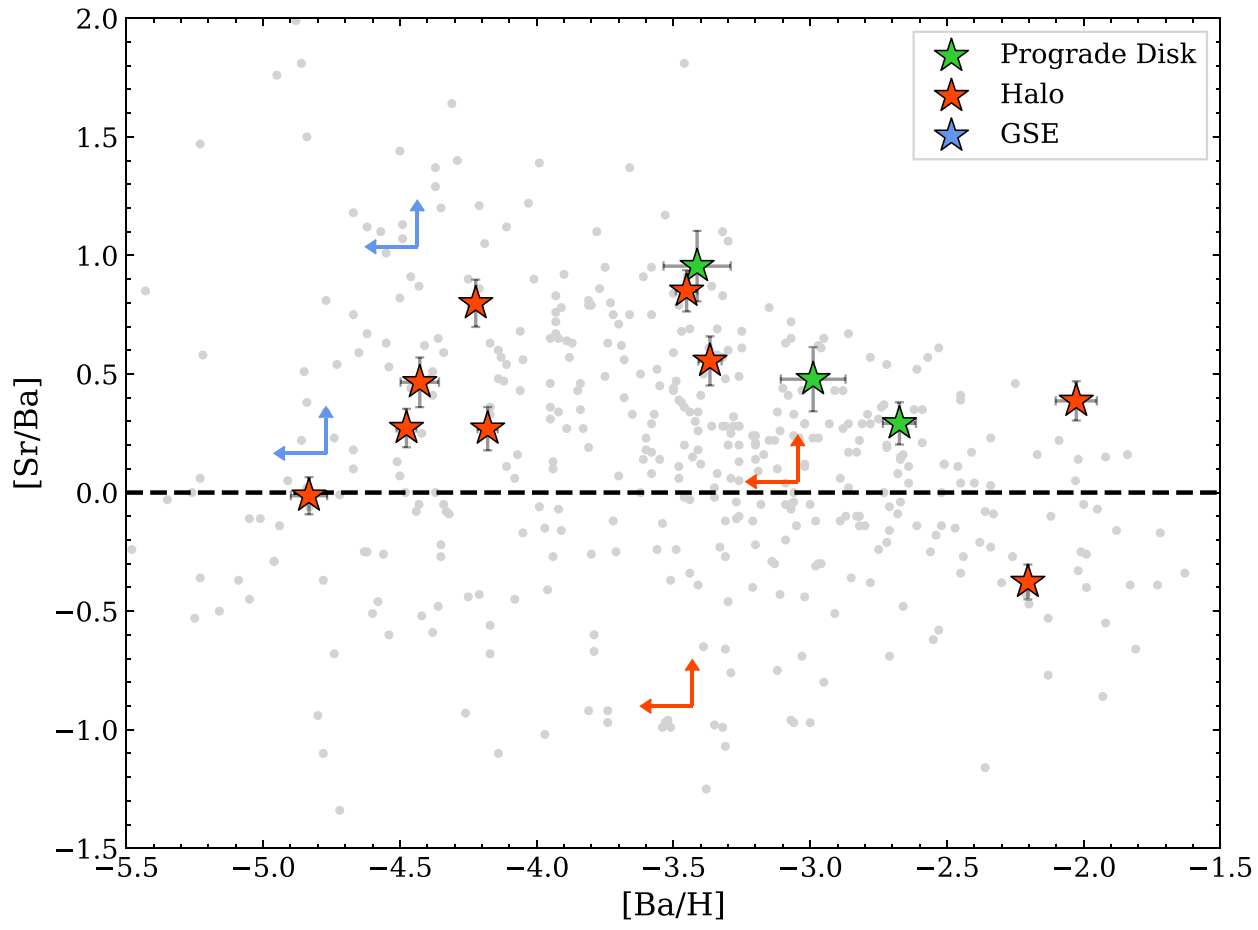

**Figure 7.** [Sr/Ba] versus [Ba/H] plot for our sample, with the values and uncertainties shown in Table 8. Stars with upper-limits in [Ba/H] (represented by leftward-facing arrows) have a corresponding lower-limit in [Sr/Ba] (represented by upward-facing arrows). The two stars with the highest [Ba/H] values refer to our r-process stars (r-I, r-II).

**Table 8.** [Ba/H] and [Sr/Ba] values for our sample. Those with an upper-limit in [Ba/H] have a corresponding lower-limit in [Sr/Ba].

| Star | [Ba/H] | [Sr/Ba] |
|---|---|---|
| ra_0103-7050_s163 | $-2.67 \pm 0.06$ | $0.29 \pm 0.09$ |
| ra_0834-5220_s316 | $-2.99 \pm 0.12$ | $0.48 \pm 0.14$ |
| ra_1604-2712_s24 | $-3.37 \pm 0.04$ | $0.56 \pm 0.10$ |
| ra_1604-2712_s292 | $< -3.43$ | $> -0.90$ |
| ra_1624-2150_s278 | $< -3.05$ | $>0.05$ |
| ra_1633-2814_s130 | $< -4.77$ | $>0.17$ |
| ra_1633-2814_s284 | $< -4.44$ | $>1.04$ |
| ra_1648-0653_s38 | $-3.41 \pm 0.12$ | $0.95 \pm 0.15$ |
| ra_1656-1433_s143 | $-2.03 \pm 0.08$ | $0.39 \pm 0.08$ |
| ra_1658-2454_s22 | $-4.22 \pm 0.03$ | $0.80 \pm 0.10$ |
| ra_1659-2154_s114 | $-4.83 \pm 0.07$ | $-0.01 \pm 0.08$ |
| ra_1709-2130_s102 | $-4.43 \pm 0.07$ | $0.47 \pm 0.10$ |
| ra_1752-4300_s214 | $-3.45 \pm 0.04$ | $0.85 \pm 0.09$ |
| ra_1752-4300_s269 | $-4.48 \pm 0.04$ | $0.27 \pm 0.08$ |
| ra_1752-4300_s6 | $-4.18 \pm 0.04$ | $0.27 \pm 0.09$ |
| ra_1853-3255_s45 | $-2.20 \pm 0.02$ | $-0.38 \pm 0.07$ |

*Note.* The full table is available in machine-readable format in the electronic version of the paper.

with theories like the ‘early’ s-process and the i-process (both alongside Sr), and weak s-process (C. M. Raiteri et al. 1991a,b) being proposed. From our sample, we have two stars with high [Ba/Fe] abundances: ra_1853-3255_s45 at $[\mathrm{Ba/Fe}] = 0.73 \pm 0.02$, and ra_1656-1433_s143 at $[\mathrm{Ba/Fe}] = 0.83 \pm 0.08$, both of which also have high abundances of [Eu/Fe]. We also have six stars with $[\mathrm{Ba/Fe}] < -1.0$: five of them detections, and one with a deep non-detection.

For Eu, we only have two stars out of 16 with detections from the Eu II lines 4129.73 and 4205.04 Å, both of them halo stars. The two Ba-rich stars, ra_1853-3255_s45 and ra_1656-1433_s143, are also enhanced in Eu, having $[\mathrm{Eu/Fe}] = 0.92 \pm 0.04$ and $[\mathrm{Eu/Fe}] = 1.32 \pm 0.03$, respectively. We adopt the definitions for two levels of r-process enhanced stars from N. Christlieb et al. (2004): r-I (moderately r-process enhanced; $0.3 \leq [\mathrm{Eu/Fe}] \leq 1.0$ and $[\mathrm{Ba/Eu}] < 0$) and r-II (strongly r-process enhanced; $[\mathrm{Eu/Fe}] > 1.0$ and $[\mathrm{Ba/Eu}] < 0$). Based on this, star ra_1853-3255_s45 is r-I ($[\mathrm{Eu/Fe}] = 0.92 \pm 0.04$ and $[\mathrm{Ba/Eu}] = -0.18 \pm 0.05$) and star ra_1656-1433_s143 is r-II ($[\mathrm{Eu/Fe}] = 1.32 \pm 0.03$ and $[\mathrm{Ba/Eu}] = -0.49 \pm 0.08$).

We show our [Sr/Ba] versus [Ba/H] plot in Fig. 7 and in Table 8. According to T. M. Sitnova et al. (2025) (using the models from F. Rizzuti et al. 2025), r-process yields $[\mathrm{Ba/Eu}] = -0.87$ and $[\mathrm{Sr/Ba}] = -0.31$. Our r-I star, ra_1853-3255_s45, has similar [Sr/Ba], but has [Ba/Eu] greater by ∼0.7 dex, giving us one indicator saying it has pure r-process. For our r-II star, ra_1656-1433_s143, [Ba/Eu] is greater by 0.3 dex, and [Sr/Ba] larger by 0.6 dex. This suggests a mixture of r-process and s-process contribution, unusual for an r-II star. This will be discussed further in Section 5.2.

## 5 DISCUSSION

The outcomes of our chemical abundance analysis are shown in Fig. 2. The panels show that despite our comparatively low-resolution spectra (5400 and 8900 for UVB and VIS, respectively), our results are in excellent agreement with those in the literature that are based on high-resolution spectra (lowest $R \sim 22\,000$). This is particularly the case for the $\alpha$- and iron-peak elements. Such success has been achieved before using X-Shooter, with E. Caffau et al. (2011b, 2013) measuring reliable chemical abundances for small samples of EMP stars. This instrument was also used in E. Caffau et al. (2011a, 2012) to identify the ‘Caffau’ star, a prograde ultra metal-poor disc star with $[\mathrm{Fe/H}] = -4.89 \pm 0.10$ (F. Sestito et al. 2019). Our work shows that this success can be extended to a larger number of elements: whilst E. Caffau et al. (2011b) measured at most 12 abundances for two stars, we were able to extend this to 16 elements across our whole sample of 16 stars. These results demonstrate that reliable abundances can be measured in stars as faint as G ≈ 17.5 mag, thereby drastically expanding the pool of metal-poor stars that can be studied in detail.

In our sample, two stars are GSE candidates, three in the prograde disc, and the rest in the halo. Of particular interest are the GSE stars: our most metal-poor star in the sample, ra_1633-2814_s284 at $[\mathrm{Fe/H}] = -3.89 \pm 0.07$, is a GSE member, and the second GSE member, ra_1633-2814_s130 is the third most metal-poor at $[\mathrm{Fe/H}] = -3.35 \pm 0.07$ (and is furthermore a Li- and Na-enhanced NEMP star, see Section 5.1). EMP stars in the GSE are rare, with the metallicity tail typically ending around $[\mathrm{Fe/H}] \approx -3.0$ (e.g. D. K. Feuillet et al. 2020; R. P. Naidu et al. 2020; P. Bonifacio et al. 2021; G. Cordoni et al. 2021).

However, recent studies have been starting to find more EMP stars in the GSE. One study by R. Zhang et al. (2024) identified five GSE stars with $[\mathrm{Fe/H}] < -3.5$, two of which have $[\mathrm{Fe/H}] < -4.0$. The study by V. M. Placco et al. (2025) found a GSE star with $[\mathrm{Fe/H}] = -4.12$. Together with our results, the number of known GSE stars with $[\mathrm{Fe/H}] < -3.5$ is now eight. These stars provide an opportunity to study the early stages of the chemical evolution of the the GSE.

In Fig. 7, we show [Sr/Ba] versus [Ba/H] for our sample. D. Yong et al. (2013) showed that the intrinsic spread of [Sr/Ba] increases with decreasing [Ba/H] when $[\mathrm{Ba/H}] < -2.5$. The increasing spread in [Sr/Ba] at lower metallicities is suggested to be due to variations in spinstar contributions, a possible pathway for the enrichment of s-process elements for stars with $[\mathrm{Fe/H}] < -2.0$ (G. Cescutti & C. Chiappini 2014). This was seen in the Phoenix stream from A. R. Casey et al. (2021), where the star-to-star variations of Sr were ascribed to varying spinstar contributions.

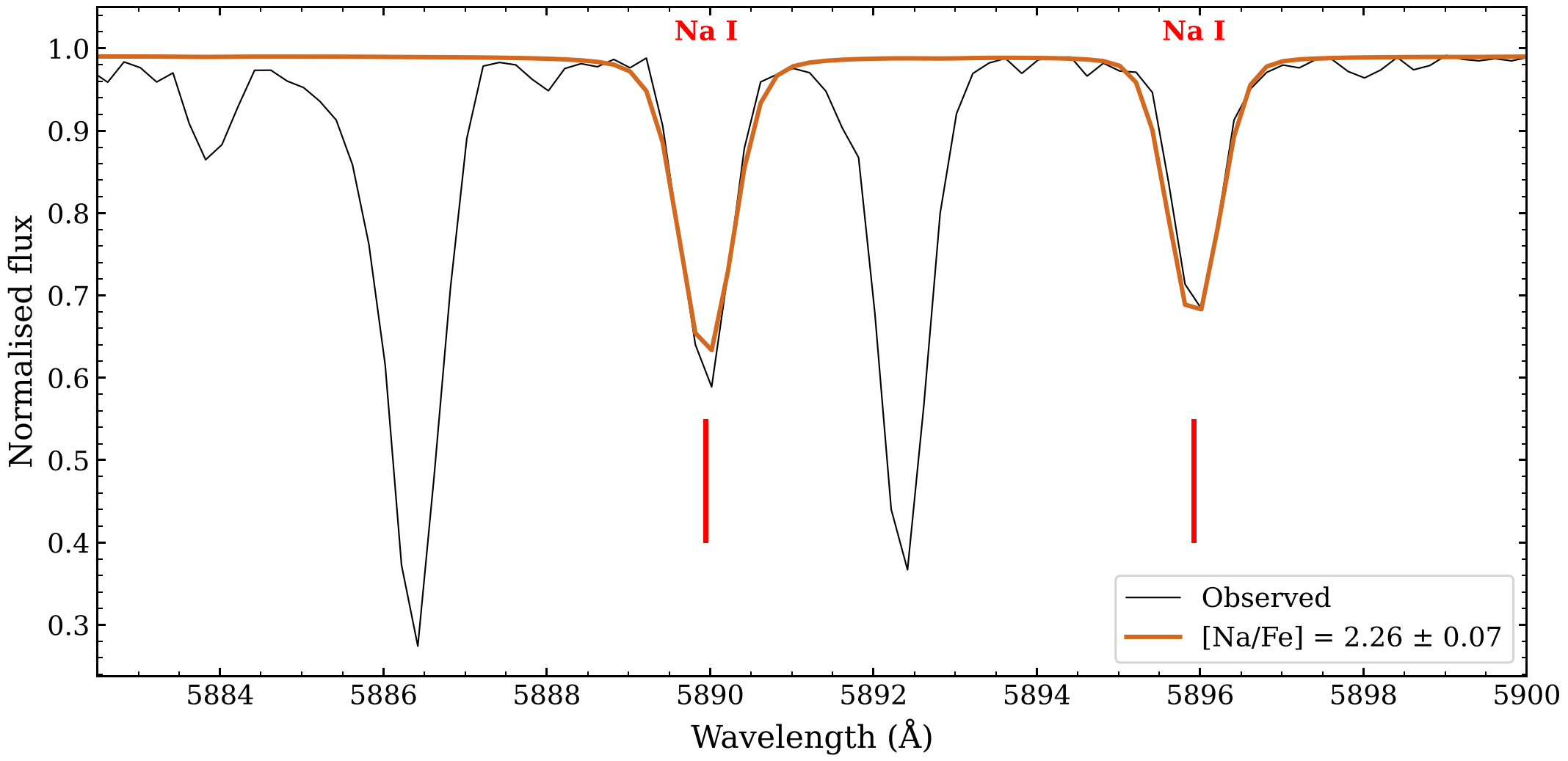

**Figure 8.** Spectral fits to the Na I 5889.95 and 5895.92 Å lines for star ra_1633-2814_s130. The thin line is the observed spectra, with the thick line representing the fit. The two strong features either side of 5889.95 Å are interstellar Na I absorption lines.

Among our sample, we have five stars with [Sr/Ba] $>0.5$ (plus another two whose error bars permit [Sr/Ba] $>0.5$). Four of these have [Sr/Ba] $>0.8$. Taking into account errors, two of the stars with [Sr/Ba] $>0.5$ belong to the prograde disc: star ra_1648-0653_s38 at [Sr/Ba] $=0.95\pm0.15$, and star ra_0834-5220_s316 at [Sr/Ba] $=0.48\pm0.14$.

Interestingly, our r-I star, ra_1853-3255_s45, has the lowest [Sr/Ba] abundance among our sample at [Sr/Ba] $=-0.38\pm0.07$, alongside having [Ba/H] $=-2.20\pm0.02$ and [Ba/Eu] $=-0.18\pm0.05$. This is in contrast with the r-II star, ra_1656-1433_s143, which has moderate abundance enhancement of [Sr/Ba] $=0.39\pm0.08$, alongside [Ba/H] $=-2.03\pm0.08$ and [Ba/Eu] $=-0.49\pm0.08$. This differs from the results seen in T. M. Sitnova et al. (2025), where r-II stars have lower [Sr/Ba] ratios than r-I stars. However, the literature sample analysed by P. Saraf et al. (2023), indicates that r-II stars reach values as high as [Sr/Ba] $\approx 1.0$ for metallicities similar to that of our star.

Among our sample we have the following chemically peculiar stars: ra_1633-2814_s130 (NEMP with large Na abundances, and as will be discussed, it is also Li-enhanced), ra_1656-1433_s143 (high Sr abundances; r-process II; mixed r-process and s-process enrichment) and ra_1658-2454_s22 (N-depleted). These will now be discussed below in detail.

### 5.1 Na- and Li-enhanced NEMP GSE star ra_1633-2814_s130

The chemical abundance pattern for the GSE star ra_1633-2814_s130 ([Fe/H] $=-3.35\pm0.07$) is characterized by [C/Fe] $=0.26\pm0.08$, [N/Fe] $=1.60\pm0.10$, and [Na/Fe] $=2.26\pm0.07$, making it an extremely Na-rich NEMP star. The fits to the Na I 5889.95 and 5895.92 Å lines for this star are shown in Fig. 8. To verify the strong Na I presence, we also examined the 8183.26 and 8194.82 Å lines using synthetic spectra at the given [Na/Fe] abundance, as seen in Fig. 9. Using star ra_1752-4300_s6 as a reference star (shifted and telluric lines scaled to match ra_1633-2814_s130), it is clear that despite the tellurics present, ra_1633-2814_s130 indeed has strong Na absorption. The strong enhancement in Na is unusual, and below, we attempt to explore the various possibilities that could explain this. We accompany the discussion with the impact of NLTE departures on our measured abundance. Besides this, we will also discuss the surprising presence of Li at $A(\mathrm{Li})_{\mathrm{3DNLTE}}=1.90\pm0.08$, with the fits shown in Fig. 10.

As discussed in Section 4.2.2, LTE Na abundances derived from the D-lines are subject to modification as a consequence of NLTE effects. At sufficiently large LTE [Na/Fe] values, the NLTE correction likely saturates. While numerous 1D NLTE abundance correction grids exist (e.g. K. Lind et al. 2011, 2022; S. A. Alexeeva et al. 2014; L. Mashonkina et al. 2023), our star ([Na/Fe] $=2.26\pm0.07$) lies outside their coverage, ending at [Na/Fe] $=1.0$. At this [Na/Fe] value, the NLTE correction for $T_{\mathrm{eff}}=5000\pm200$ K, $\log g=1.85\pm0.78$, [Fe/H] $=-3.35\pm0.07$, and $v_{\mathrm{mic}}=2.04\pm0.65$ km s$^{-1}$ is $-0.55$ dex, where the value is the average of the individual corrections for the Na D lines. The star is thus still significantly enhanced in Na. It is unlikely that the NLTE correction at the observed LTE abundance of [Na/Fe] $=2.26\pm0.07$ is substantially different from $-0.55$ dex, with it likely saturating after [Na/Fe] $=1.0$ based on the models used in K. Lind et al. (2022).

There are two stars in the literature that have similar abundances to what we see in ra_1633-2814_s130. The first is the LMC star SMSS DR3 497 519 424 from W. S. Oh et al. (2024), which has a metallicity ([Fe/H] $=-3.13$) similar to our star. It was identified as a NEMP with [N/Fe] $=1.70\pm0.11$ and [C/Fe] $=0.63\pm0.14$. It also shows enhancements in Na and Al ([Na/Fe] $=1.25\pm0.15$ and [Al/Fe] $=-0.10\pm0.21$) but, like our star, the Mg abundance is normal ([Mg/Fe] $=0.43\pm0.05$). The LMC also has low Sr and Ba at [Sr/Fe] $=-0.59\pm0.17$ and [Ba/Fe] $=-0.70\pm0.15$, inconsistent with our star. The authors suggested that the reason for these abundance patterns was due to rotation in the progenitor star with little s-process enrichment. If we discard Al and assume that the process that is making N and Na, but not Mg, then the LMC star is a reasonable match to our star. The second is star SMSS J215805.81−651327.2 from R. Cayrel et al. (2004) and H. R. Jacobson et al. (2015), also having comparable metallicities to ours ([Fe/H] $=-3.41$). This star has similar abundances to what we find, having [Na/Fe] $=1.93$, [Mg/Fe] $=0.42$, [Al/Fe] $=-1.0$, [Sr/Fe] = −0.56, and [Ba/Fe] = −0.97 (from H. R. Jacobson et al. 2015). The only discrepancy is in N and C, being determined at [C/Fe] = 0.27 and [N/Fe] = 0.71 (from M. Spite et al. (2005)). No explanation was given by the authors, though future work on ra_1633-2814_s130 could benefit from further analysis on SMSS J215805.81−651327.2.

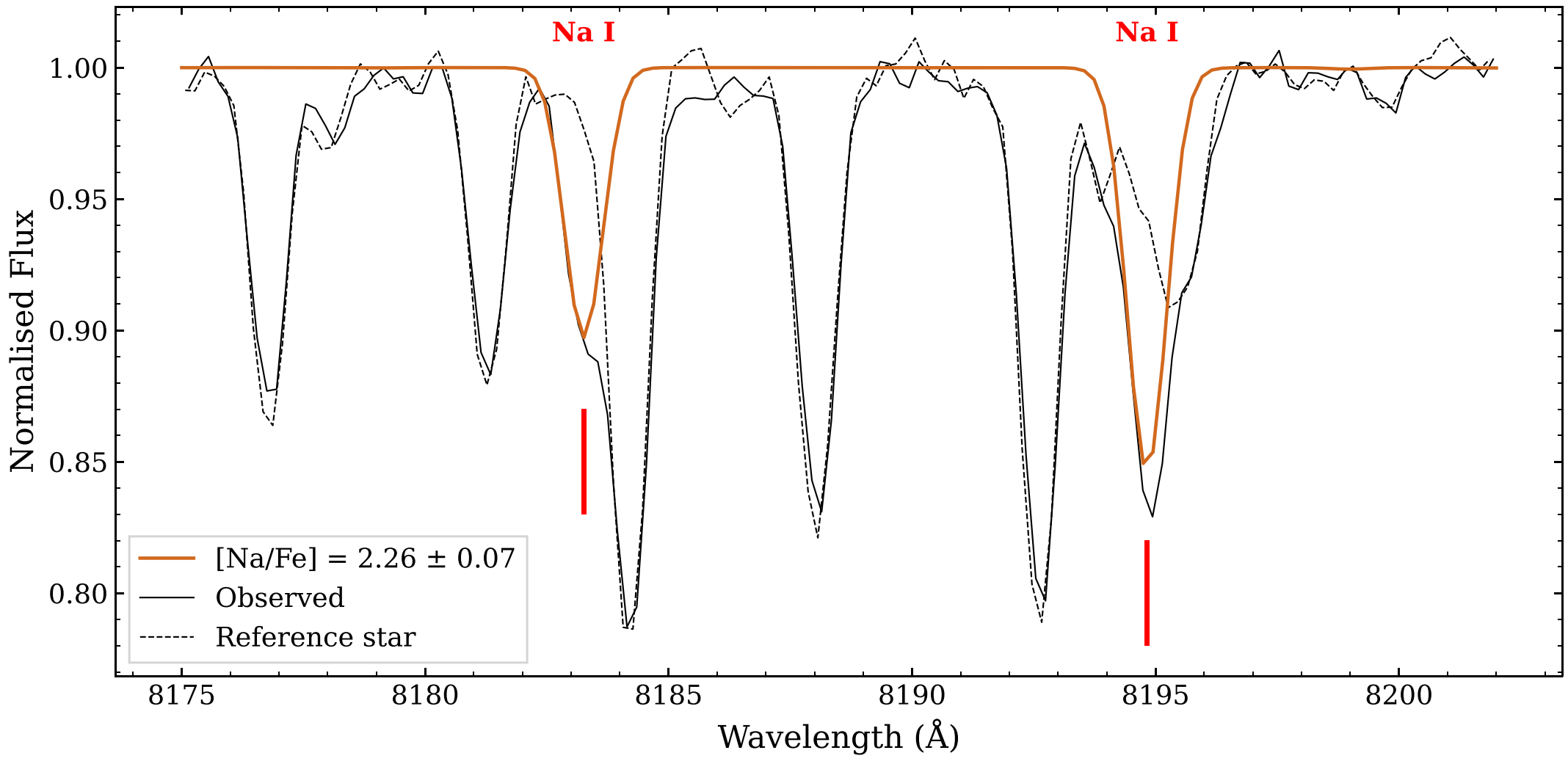

**Figure 9.** Synthetic spectra of the Na I 8183.26 and 8194.82 Å lines for star ra_1633-2814_s130. The thin line is the observed spectrum (which is dominated by telluric lines), the thick line is the synthetic spectra calculated at [Na/Fe] = 2.26 ± 0.07, and the dotted line is star ra_1752-4300_s6, chosen as our reference. The tellurics in the reference star have been shifted and scaled to match ra_1633-2814_s130. Star ra_1752-4300_s6 was selected because of its relatively low Na I abundance at [Na/Fe] = 0.04 ± 0.05.

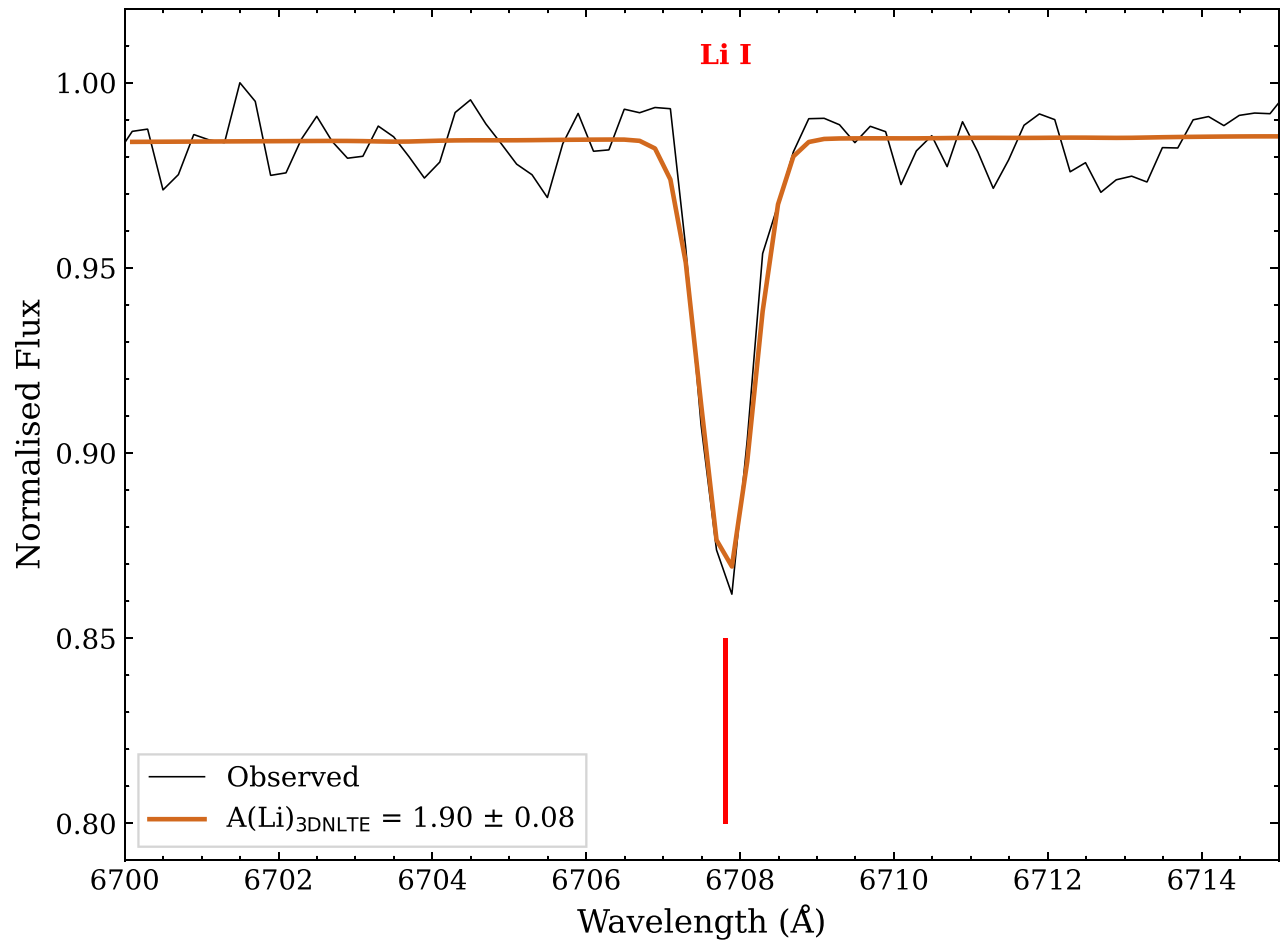

**Figure 10.** Spectral fit to the Li I 6709.66 Å lines for star ra_1633-2814_s130. The 1D LTE Li abundance was corrected to 3D NLTE using the `Breidablik` code (E. X. Wang et al. 2021, 2024); the correction was $\Delta_{3\mathrm{DNTLE}} = -0.08$.

With the lack of analogues in the literature, we considered other possibilities to describe this star. One such possibility is that this star was an escapee from a globular cluster. Globular clusters (GCs) are known to have two distinct stellar populations separated by chemical compositions: a first population (1P), having similar chemistry to halo field stars, and a second population (2P), having enhancements in He, N, Na, and Al, alongside having deficiencies in C, O, and Mg (R. P. Kraft 1994; N. Bastian & C. Lardo 2018; R. Gratton et al. 2019; A. P. Milone & A. F. Marino 2022). Star ra_1633-2814_s130 is enhanced in N and Na, and within uncertainty, is also depleted in C. 1P/2P Al variations are seen in GCs at low metallicities (e.g. D. M. Nataf et al. 2019), though GCs at metallicities of our star is very rare, making the connection difficult. Given this, the 2P connection is unlikely for ra_1633-2814_s130.

Intriguingly, this star has a large abundance of Li, as shown in Fig. 10, with a 3D NLTE abundance of $A(\mathrm{Li})_{3\mathrm{DNLTE}} = 1.90 \pm 0.08$. At $T_{\mathrm{eff}} = 5000 \pm 200$ K and $\log g = 1.85 \pm 0.78$, this star falls within the RGB plateau, a region found in the lower RGB after the first dredge-up episode, but before evolutionary mixing that destroys Li (A. Mucciarelli et al. 2022, see their fig. 1). Despite the large error bars on log g, our star has a formal upper-limit of $\log g < 3.23 \pm 0.02$ directly from its parallax, which firmly places it on the RGB plateau. Stars here typically have $A(\mathrm{Li})_{3\mathrm{DNLTE}}$ values in the range 0.87–1.23 dex. Therefore on average, our star possibly has a Li enhancement of 0.9 dex. More work is needed to confirm the process that could cause this strong Li enhancement, alongside the enhancements seen in both Na and N.

Given the large errors on log g, it might also be possible that this star is instead a horizontal branch star. These objects typically have a log g between 2.5 and 2.8 dex, and a $T_{\mathrm{eff}}$ ranging from 4600 to 5000 K (e.g. L. Girardi 2016), which overlaps the parameter space of our star. High Li abundances are known in horizontal branch stars (e.g. M. Spite & F. Spite 1982; G. R. Ruchti et al. 2011; H. Li et al. 2018; A. R. Casey et al. 2019; H.-L. Yan et al. 2021; A. Susmitha, A. Mallick & B. E. Reddy 2024), with typical abundances around A(Li) ≈ 2.0 dex. If our star does lie on the horizontal branch, then the large abundance of Li seen in this star may arise from internal gravity wave induced mixing transferring large amounts of Be from the hydrogen-burning shell to the cooler envelopes, where it is converted into Li (e.g. F. W. Wu et al. 2025). However, no studied horizontal branch stars with high Li show strong enhancements in both Na and N (e.g. G. R. Ruchti et al. 2011; A. Susmitha et al. 2024).

With strong enhancements in N ([N/Fe] = 1.60 ± 0.10), Na ([Na/Fe] = 2.26 ± 0.07), Li ($A(\mathrm{Li})_{3\mathrm{DNLTE}} = 1.90 \pm 0.08$), and all other measured elements are consistent with those seen in halo stars (see panel in Fig. 3), we have a type of star with only two possible analogues: SMSS DR3 497519424 and SMSS J215805.81−651327.2, with the latter also having a Li abundance of A(Li)$_{\rm NLTE}$ = 0.90 (from H. R. Jacobson et al. 2015). At this stage, we cannot determine the origin of the unusual abundance pattern in this star.

### 5.2 Neutron-capture enhanced star ra_1656-1433_s143

The chemical abundance pattern for the halo star ra_1656-1433_s143 ([Fe/H] $= -2.86 \pm 0.06$) is characterized by its neutron-capture abundance enhancement ([Ba/Fe] $= 0.83 \pm 0.08$, [Eu/Fe] $= 1.32 \pm 0.03$, [Sr/Fe] $= 1.22 \pm 0.03$), classifying it as r-II. Of particular interest is surprisingly high ratio [Sr/Ba] $= 0.39 \pm 0.08$, at an abundance of [Ba/H] $= -2.03 \pm 0.08$. Most r-II stars have negative [Sr/Ba] values, but our star has a positive value. We discuss the implications of this result below.

Several production sites for r-II stars have been proposed in the literature, such as from core-collapse supernovae (N. Nishimura, T. Takiwaki & F.-K. Thielemann 2015; T. Tsujimoto & N. Nishimura 2015; P. Mösta et al. 2018) and neutron star mergers (A. P. Ji et al. 2016; D. Kasen et al. 2017; E. Pian et al. 2017), though the majority of r-II stars agree with the latter (e.g. A. Bandyopadhyay et al. 2024). Finding these stars were initially rare, with success rates being around ∼3 per cent (P. S. Barklem et al. 2005; A. Frebel 2018; D. Yong et al. 2021a), but recent studies like A. Bandyopadhyay et al. (2024) have been more successful at ∼10 per cent. From our small sample, our discovery rate is 6.25 per cent.

As stated in T. M. Sitnova et al. (2025), r-II stars typically have [Ba/Eu] $= -0.87$ and [Sr/Ba] $= -0.31$, indicating a lack of any s-process contribution. In contrast, r-I stars show greater scatter in [Sr/Ba], with values either positive or negative, a result of more moderate r-process enhancement mixed with other nucleosynthetic events. This is seen in the literature: all four r-II stars from D. Yong et al. (2021b) have negative [Sr/Ba] values ranging from −0.10 to −0.38; the two r-II stars studied in P. Saraf et al. (2023) have [Sr/Ba] ≈ 0, whilst all three r-II stars in T. M. Sitnova et al. (2025) have [Sr/Ba] < −0.25. A handful of r-II stars do exist in the literature with positive values, reaching to [Sr/Ba] ≈ 1.0 for similar metallicities, as shown in fig. 16 of P. Saraf et al. (2023) (with references therein).

Our star, with [Ba/Eu] $= -0.49 \pm 0.08$ and [Sr/Ba] $= 0.38 \pm 0.07$, matches these latter stars in having uncommon [Ba/Eu] and [Sr/Ba] abundances for r-II stars. This suggests our star is not purely r-process enhanced, rather a mixture of r-process (creating Ba) and s-process enhancement (creating Sr). This situation suggests that alongside the normal r-II formation channels (core-collapse supernova and neutron star mergers), this star also underwent s-process enhancement from a source that can operate at low metallicities such as spinstars (e.g. G. Cescutti et al. 2013; P. Banerjee et al. 2018). Abundance determinations for additional s- and r-process elements would be beneficial to constrain the nucleosynthetic processes generating the observed abundances in this star.

### 5.3 N-depleted halo star ra_1658-2454_s22

The chemical abundances for the halo star ra_1658-2454_s22 ([Fe/H] $= -2.90 \pm 0.08$) is characterized by its very low N upper limit of [N/Fe] < −1.11, alongside having C depleted at [C/Fe] $= -0.33 \pm 0.08$ with zero evolutionary correction. Every other abundance measurement is consistent with the comparison literature (see panel in Fig. 3). We show the NH region for this in the top right panel of Fig. B2, including synthetic spectra at [N/Fe] $= -1.11$ (dark blue) and 0.50 (dash light blue). Here, we will discuss the implications of this result, alongside any reasons for N being this depleted.

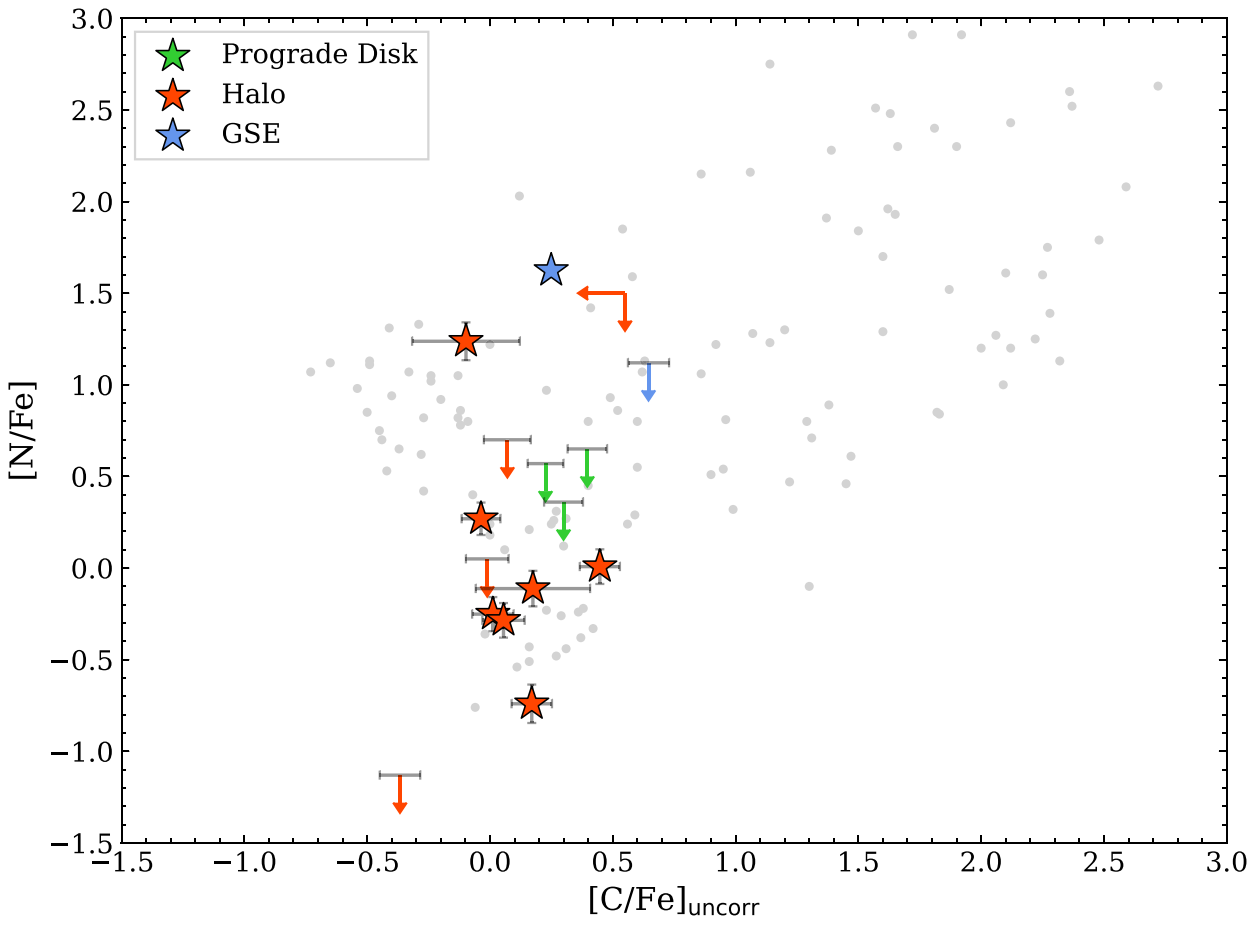

**Figure 11.** Uncorrected [C/Fe] versus [N/Fe] abundances for our sample and of the comparison literature. The N-depleted star, ra_1658-2454_s22, is the outlying upper limit located in the lower left of the plot. Note that the comparison sample (grey) has not had its evolutionary corrections removed.

In Fig. 11, we show our uncorrected [C/Fe] versus [N/Fe] for both our sample and of the comparison literature. It is clear that star ra_1658-2454_s22 is an outlier in both data sets, with the two closest stars belonging to ra_1853-3255_s45 (with [N/Fe] $= -0.66 \pm 0.10$ and [C/Fe]$_{\rm uncorr}$ $= 0.20 \pm 0.08$), alongside CS 29516–024 from M. Spite et al. (2005) and D. Yong et al. (2013) (with [N/Fe] $= -0.76 \pm 0.10$ and [C/Fe]$_{\rm uncorr}$ = 0.69). The latter star has low $\log g$ (1.04 dex), thus having a high evolutionary correction of +0.75. This means that C has depleted a substantial amount, being converted to N in the process. Given N for CS 29516–024 is already low, it is possible that CS 29516–024 would have an even lower abundance, possibly similar to our N-depleted star, which is yet to undergo significant evolutionary mixing.

Given that ra_1658-2454_s22 has 'normal' abundances for every other element besides C and N, whatever process(es) took place, only C and N were altered. One suggestion came from M. Spite et al. (2005), where they say that a depleted N and C star may have undergone Type II supernova enrichment to provide the 'normal' $\alpha$-abundances, but occurred before any AGB enrichment to give the C and N abundances. With the lack of s-process enrichment due to low Sr and Ba abundances ([Sr/Fe] $= -0.53 \pm 0.10$ and [Ba/Fe] $= -1.32 \pm 0.03$), which is otherwise enriched by AGB stars, this idea is plausible. A follow-up high-resolution study to determine additional abundances in this star, particularly that for O, would be worthwhile.

## 6 SUMMARY AND CONCLUSION

In this work, we have performed follow-up observations on 16 metal-poor star candidates selected from B. D. C. Lowe et al. (2025), using the medium-resolution X-Shooter spectrograph. After taking advantage of the continuum normalization SUPP-NET code (T. Różański et al. 2022), we used the 1D LTE code KORG to rederive metallicities and measure 16 elemental abundances (C (CH), N (NH), Na I, Mg I, Al I, Si I, Ca II, Sc II, Ti II, Cr I, Mn I, Co I, Ni I, Sr II, Ba II, and Eu II) for our sample. From our metallicity results, we have identified six EMPs, with the lowest metallicity at $[\mathrm{Fe/H}] = -3.89 \pm 0.06$. Our metallicities are in excellent agreement with B. D. C. Lowe et al. (2025) (Fig. 1), with this work's abundances higher, on average, by $\Delta\,[\mathrm{Fe/H}] = -0.13$ dex.

The chemical abundances of our stars were compared with high-resolution literature (D. Yong et al. 2013, 2021a; H. R. Jacobson et al. 2015; A. F. Marino et al. 2019; Fig. 2), and we see excellent agreement across the 16 elements. We show that even on medium-resolution instruments like X-Shooter, we can measure a large number of elements on stars considerably fainter than the typical stars analysed in the high-resolution studies

Among our sample, we have two C-rich and one C-poor star (Fig. 4), one probable and one possible NEMP stars (Fig. 5), and two star with $[\alpha/\mathrm{Fe}] > 0.4$ (Fig. 6). When looking at the chemical abundance patterns for our sample (Fig. 3), we identified three peculiar stars. The first: ra_1633-2814_s130, a NEMP star that has unusually strong enhancement in Na ($[\mathrm{Na/Fe}] = 2.26 \pm 0.07$, with NLTE corrections unlikely to change its high Na abundance; Figs 8 and 9), and Li ($A(\mathrm{Li})_{3\mathrm{DNLTE}} = 1.90 \pm 0.08$; Fig. 10), but none in Al or in any neutron capture elements. With the presence of Li, this star either belongs to the RGB plateau (with at most $\sim$1.0 dex Li enhancement via an unknown process), or from the red clump phase. With high N, Li, and Na, but otherwise 'normal' halo abundance ratios, particularly with Al and Mg, the origin of the abundance patterns in this star remains a mystery.

The second peculiar star: ra_1656-1433_s143, categorized as r-II, has strong Sr ($[\mathrm{Sr/Fe}] = 1.22 \pm 0.03$) and positive $[\mathrm{Sr/Ba}] = 0.39 \pm 0.08$ abundances (Fig. 7). Generally, r-II stars have negative [Sr/Ba] abundances, reflecting a lack of s-process enrichment, but the observed values here likely indicate a mixture of both s-process and r-process enrichment. This suggests that ra_1656-1433_s143 was potentially formed from material enriched in both r-process material (via binary neutron star merger or core-collapse supernovae) and s-process enriched material, potentially from spinstars.

The third star: ra_1658-2454_s22, has a very low upper limit on N ($[\mathrm{N/Fe}] < -1.11$), low C with minimal evolutionary mixing corrections ($[\mathrm{C/Fe}] = -0.33 \pm 0.08$ with $\Delta\,[\mathrm{C/Fe}] = 0.00$; Fig. 11), but otherwise 'normal' [X/Fe] abundances. This is consistent with a star that underwent Type II supernova enrichment (producing the [X/Fe] abundances), but occurring before massive AGB enrichment that would generally provide the C and N abundances.

Finally, we reveal that among the six EMPs in this sample: two are GSE candidates at $[\mathrm{Fe/H}] = -3.89 \pm 0.07$ for star ra_1633-2814_s284 (most metal-poor in our sample), and at $[\mathrm{Fe/H}] = -3.35 \pm 0.07$ for the NEMP star ra_1633-2814_s130 (third most metal-poor). EMP stars in the GSE are uncommon, with only eight known at $[\mathrm{Fe/H}] < -3.5$.

## ACKNOWLEDGEMENTS

This paper includes data gathered with the 8 m VLT located at Cerro Paranal, Chile, and is based on observations collected at the European Southern Observatory under ESO programme 113.26N5.001.

We would like to express our warmest gratitude to Dr. Adam Wheeler, the creator of the 1D LTE code KORG, who provided valuable guidance on using the code for metal-poor stars.

This work was supported by computational resources provided by the Australian Government through the National Computational Infrastructure (NCI) under the National Computational Merit Allocation Scheme (project y89). TN acknowledges support from the Knut and Alice Wallenberg Foundation.

We also thank the anonymous referee for their insightful comments and suggestions, which have led to improvements in the manuscript.

## DATA AVAILABILITY

The data used in this study are available in the ESO archive (https://archive.eso.org/eso/eso_archive_main.html) under programme ID 113.26N5.001. Our co-added spectra are available upon request.

## REFERENCES

Aguado D. S., Prieto C. A., Hernández J. I. G., Rebolo R., 2018, ApJ, 854, L34
Alexeeva S. A., Pakhomov Y. V., Mashonkina L. I., 2014, Astron. Lett., 40, 406
Aoki W., Beers T. C., Christlieb N., Norris J. E., Ryan S. G., Tsangarides S., 2007, ApJ, 655, 492
Arcones A., Bliss J., 2014, J. Phys. G Nucl. Phys., 41, 044005
Arentsen A. et al., 2020, MNRAS, 491, L11
Bandyopadhyay A. et al., 2024, ApJS, 274, 39
Banerjee P., Qian Y.-Z., Heger A., 2018, ApJ, 865, 120
Barklem P. S. et al., 2005, A&A, 439, 129
Bastian N., Lardo C., 2018, ARA&A, 56, 83
Beers T. C., Christlieb N., 2005, ARA&A, 43, 531
Bellazzini M., Massari D., Ceccarelli E., Mucciarelli A., Bragaglia A., Riello M., Angeli F. D., Montegriffo P., 2024, A&A, 683, A136
Belokurov V., Erkal D., Evans N. W., Koposov S. E., Deason A. J., 2018, MNRAS, 478, 611
Belokurov V., Sanders J. L., Fattahi A., Smith M. C., Deason A. J., Evans N. W., Grand R. J. J., 2020, MNRAS, 494, 3880
Bergemann M., Cescutti G., 2010, A&A, 522, A9
Bessell M. S., Norris J., 1984, ApJ, 285, 622
Bessell M. S. et al., 2015, ApJ, 806, L16
Bonifacio P. et al., 2021, A&A, 651, A79
Buder S. et al., 2025, Publ. Astron. Soc. Aust., 42, e051
Caffau E. et al., 2011a, Nature, 477, 67
Caffau E. et al., 2011b, A&A, 534, A4
Caffau E. et al., 2012, A&A, 542, A51
Caffau E. et al., 2013, A&A, 560, A15
Caffau E. et al., 2024, A&A, 691, A245
Cai S. et al., 2025, ApJ, 993, L52
Casagrande L., Wolf C., Mackey A. D., Nordlander T., Yong D., Bessell M., 2019, MNRAS, 482, 2770
Casagrande L. et al., 2021, MNRAS, 507, 2684
Casey A. R. et al., 2019, ApJ, 880, 125
Casey A. R. et al., 2021, ApJ, 921, 67
Cayrel R., 1988, Proc. IAU Symp. 132, The Impact of Very High S/N Spectroscopy on Stellar Physics. Kluwer, Dordrecht, p. 345, https://ui.adsabs.harvard.edu/abs/1988IAUS..132..345C
Cayrel R. et al., 2004, A&A, 416, 1117
Cescutti G., Chiappini C., 2014, A&A, 565, A51
Cescutti G., Chiappini C., Hirschi R., Meynet G., Frischknecht U., 2013, A&A, 553, A51
Chiti A., Mardini M. K., Frebel A., Daniel T., 2021, ApJ, 911, L23
Choplin A., Siess L., Goriely S., 2021, A&A, 648, A119
Choplin A., Siess L., Goriely S., Martinet S., 2024, A&A, 684, A206

Christlieb N. et al., 2002, Nature, 419, 904
Christlieb N. et al., 2004, A&A, 428, 1027
Christlieb N., Schörck T., Frebel A., Beers T. C., Wisotzki L., Reimers D., 2008, A&A, 484, 721
Cordoni G. et al., 2021, MNRAS, 503, 2539
Da Costa G. S. et al., 2019, MNRAS, 489, 5900
Dovgal A. et al., 2024, MNRAS, 527, 7810
Eichler M. et al., 2018, J. Phys. G Nucl. Phys., 45, 014001
Fernández-Alvar E. et al., 2021, MNRAS, 508, 1509
Feuillet D. K., Feltzing S., Sahlholdt C. L., Casagrande L., 2020, MNRAS, 497, 109
Frebel A., 2010, Astron. Nachr., 331, 474
Frebel A., 2018, Annu. Rev. Nucl. Particle Sci., 68, 237
Frebel A. et al., 2005, Nature, 434, 871
Fujimoto S. et al., 2025, ApJ, 989, 46
Ghosh S., Wolfe N., Fröhlich C., 2022, ApJ, 929, 43
Girardi L., 2016, ARA&A, 54, 95
Gratton R., Bragaglia A., Carretta E., D'Orazi V., Lucatello S., Sollima A., 2019, A&AR, 27, 8
Greif T. H., Johnson J. L., Klessen R. S., Bromm V., 2009, MNRAS, 399, 639
Gustafsson B., Edvardsson B., Eriksson K., Jørgensen U. G., Nordlund Å., Plez B., 2008, A&A, 486, 951
Helmi A., Babusiaux C., Koppelman H. H., Massari D., Veljanoski J., Brown A. G. A., 2018, Nature, 563, 85
Hou X., Zhao G., Li H., 2024, MNRAS, 532, 1099
Howes L. M. et al., 2015, Nature, 527, 484
Howes L. M. et al., 2016, MNRAS, 460, 884
Ishigaki M. N. et al., 2021, MNRAS, 506, 5410
Izutani N., Umeda H., Tominaga N., 2009, ApJ, 692, 1517
Jacobson H. R. et al., 2015, ApJ, 807, 171
Ji A. P., Frebel A., Chiti A., Simon J. D., 2016, Nature, 531, 610
Ji A. P. et al., 2025,, https://ui.adsabs.harvard.edu/abs/2025arXiv250921643J
Jin S. et al., 2024, MNRAS, 530, 2688
Johnson J. A., Herwig F., Beers T. C., Christlieb N., 2007, ApJ, 658, 1203
de Jong R. S. et al., 2019, The Messenger, 175, 3
Kasen D., Metzger B., Barnes J., Quataert E., Ramirez-Ruiz E., 2017, Nature, 551, 80
Keller S. C. et al., 2014, Nature, 506, 463
Kielty C. L. et al., 2021, MNRAS, 506, 1438
Klessen R. S., Glover S. C., 2023, ARA&A, 61, 165, https://doi.org/10.1146/annurev-astro-071221-053453
Kraft R. P., 1994, PASP, 106, 553
Kupka F., Piskunov N., Ryabchikova T. A., Stempels H. C., Weiss W. W., 1999, A&AS, 138, 119
Lagae C., Amarsi A. M., Díaz L. F. R., Lind K., Nordlander T., Hansen T. T., Heger A., 2023, A&A, 672, A90
Li H., Aoki W., Matsuno T., Kumar Y. B., Shi J., Suda T., Zhao G., 2018, ApJ, 852, L31
Li T. S. et al., 2022, ApJ, 928, 30
Limberg G., Souza S. O., Pérez-Villegas A., Rossi S., Perottoni H. D., Santucci R. M., 2022, ApJ, 935, 109
Limberg G., Placco V. M., Ji A. P., Yao Y., Chiti A., Mardini M. K., Frebel A., Rossi S., 2025, ApJ, 989, L18
Lin Y. et al., 2025, ApJ, 984, L43
Lind K., Asplund M., Barklem P. S., Belyaev A. K., 2011, A&A, 528, A103
Lind K. et al., 2022, A&A, 665, A33
Lowe B. D. C. et al., 2025, Publ. Astron. Soc. Aust., 42, e040
Marino A. F. et al., 2019, MNRAS, 485, 5153
Mas-Ribas L., Dijkstra M., Forero-Romero J. E., 2016, ApJ, 833, 65
Mashonkina L., Pakhomov Y., Sitnova T., Smogorzhevskii A., Jablonka P., Hill V., 2023, MNRAS, 524, 3526
Milone A. P., Marino A. F., 2022, Universe, 8, 359
Mondal C. et al., 2025, ApJ, 988, 171
Morishita T., Liu Z., Stiavelli M., Treu T., Bergamini P., Zhang Y., 2025,, https://ui.adsabs.harvard.edu/abs/2025arXiv250710521M
Mösta P., Roberts L. F., Halevi G., Ott C. D., Lippuner J., Haas R., Schnetter E., 2018, ApJ, 864, 171
Mucciarelli A., Monaco L., Bonifacio P., Salaris M., Deal M., Spite M., Richard O. A., Lallement R., 2022, A&A, 661, A153
Naidu R. P., Conroy C., Bonaca A., Johnson B. D., Ting Y.-S., Caldwell N., Zaritsky D., Cargile P. A., 2020, ApJ, 901, 48
Nakajima K. et al., 2025, preprint (arXiv:2506.11846), http://arxiv.org/abs/2506.11846
Nataf D. M. et al., 2019, AJ, 158, 14
Nishimura N., Takiwaki T., Thielemann F.-K., 2015, ApJ, 810, 109
Nordlander T., Amarsi A. M., Lind K., Asplund M., Barklem P. S., Casey A. R., Collet R., Leenaarts J., 2017, A&A, 597, A6
Nordlander T. et al., 2019, MNRAS, 488, L109
Oh S. P., Haiman Z., Rees M. J., 2001, ApJ, 553, 73
Oh W. S., Nordlander T., Da Costa G. S., Bessell M. S., Mackey A. D., 2023, MNRAS, 524, 577
Oh W. S., Nordlander T., Da Costa G. S., Bessell M. S., Mackey A. D., 2024, MNRAS, 528, 1065
Onken C. A. et al., 2019, Publ. Astron. Soc. Aust., 36, e033
Onken C. A., Wolf C., Bessell M. S., Chang S.-W., Luvaul L. C., Tonry J. L., White M. C., Da Costa G. S., 2024, Publ. Astron. Soc. Aust., 41, e061
Osorio Y., Aguado D. S., Allende Prieto C., Hubeny I., González Hernández J. I., 2022, ApJ, 928, 173
Pian E. et al., 2017, Nature, 551, 67
Piskunov N. E., Kupka F., Ryabchikova T. A., Weiss W. W., Jeffery C. S., 1995, A&AS, 112, 525
Placco V. M., Frebel A., Beers T. C., Stancliffe R. J., 2014, ApJ, 797, 21
Placco V. M. et al., 2025, ApJ, 991, 101
Qian Y. Z., Wasserburg G. J., 2007, Phys. Rep., 442, 237
Raiteri C. M., Busso M., Gallino R., Picchio G., Pulone L., 1991a, ApJ, 367, 228
Raiteri C. M., Busso M., Gallino R., Picchio G., 1991b, ApJ, 371, 665
Riaz S., Hartwig T., Latif M. A., 2022, ApJ, 937, L6
Rizzuti F., Cescutti G., Molaro P., Roberti L., Chieffi A., Limongi M., Magrini L., Matteucci F., 2025, A&A, 698, A118
Różański T., Niemczura E., Lemiesz J., Posiłek N., Różański P., 2022, A&A, 659, A199
Ruchti G. R. et al., 2011, ApJ, 743, 107
Ryabchikova T., Piskunov N., Kurucz R. L., Stempels H. C., Heiter U., Pakhomov Y., Barklem P. S., 2015, Phys. Scr., 90, 054005
Rydberg C.-E., Zackrisson E., Lundqvist P., Scott P., 2013, MNRAS, 429, 3658
Salvadori S., Ferrara A., Schneider R., Scannapieco E., Kawata D., 2010, MNRAS, 401, L5
Sana H. et al., 2024, A&A, 688, A104
Santistevan I. B., Wetzel A., Sanderson R. E., El-Badry K., Samuel J., Faucher-Giguère C.-A., 2021, MNRAS, 505, 921
Saraf P., Allende Prieto C., Sivarani T., Bandyopadhyay A., Beers T. C., Susmitha A., 2023, MNRAS, 524, 5607
Saunders W. et al., 2004, in Proc. SPIE Conf. Ser., Ground-based Instrumentation for Astronomy. SPIE, Bellingham, p. 389, https://www.spiedigitallibrary.org/conference-proceedings-of-spie/5492/0000/AAOmega-a-scientific-and-optical-overview/10.1117/12.550871.full
Scannapieco E., Schneider R., Ferrara A., 2003, ApJ, 589, 35
Schlegel D. J., Finkbeiner D. P., Davis M., 1998, ApJ, 500, 525
Schörck T. et al., 2009, A&A, 507, 817
Sestito F. et al., 2019, MNRAS, 484, 2166
Sestito F. et al., 2020, MNRAS, 497, L7
Sharp R. et al., 2006, in McLean I. S., Iye M. eds, Proc. SPIE Conf. Ser. Vol. 6269, Ground-based and Airborne Instrumentation for Astronomy. SPIE, Bellingham, p. 62690G
Simmerer J., Sneden C., Cowan J. J., Collier J., Woolf V. M., Lawler J. E., 2004, ApJ, 617, 1091
Sitnova T. M. et al., 2025, A&A, 699, A262
Spite M., Spite F., 1982, Nature, 297, 483
Spite M. et al., 2005, A&A, 430, 655
Starkenburg E. et al., 2018, MNRAS, 481, 3838
Susmitha A., Mallick A., Reddy B. E., 2024, ApJ, 966, 109

Tsujimoto T., Nishimura N., 2015, ApJ, 811, L10
Vernet J. et al., 2011, A&A, 536, A105
Vincenzo F., Spitoni E., Calura F., Matteucci F., Silva Aguirre V., Miglio A., Cescutti G., 2019, MNRAS, 487, L47
Wang E. X., Nordlander T., Asplund M., Amarsi A. M., Lind K., Zhou Y., 2021, MNRAS, 500, 2159
Wang E. X. et al., 2024, MNRAS, 528, 5394
Wheeler A. J., Abruzzo M. W., Casey A. R., Ness M. K., 2022, Astrophysics Source Code Library, record (ascl:2211.016)
Wheeler A. J., Abruzzo M. W., Casey A. R., Ness M. K., 2023, AJ, 165, 68
Wolf C. et al., 2018, Publ. Astron. Soc. Aust., 35, e010
Woosley S. E., Hoffman R. D., 1992, ApJ, 395, 202
Wu F. W. et al., 2025, A&A, 693, A138
Yan H.-L. et al., 2021, Nat. Astron., 5, 86
Yong D. et al., 2013, ApJ, 762, 26
Yong D. et al., 2021a, MNRAS, 507, 4102
Yong D. et al., 2021b, Nature, 595, 223
Zackrisson E., Rydberg C.-E., Schaerer D., Östlin G., Tuli M., 2011, ApJ, 740, 13
Zackrisson E. et al., 2012, MNRAS, 427, 2212
Zhang R. et al., 2024, ApJ, 966, 174
Ďurovčíková D. et al., 2025, ApJ, 987, L33

## SUPPORTING INFORMATION

Tables in the manuscript (Tables 1–8) and Supplementary Material (Tables S1–S4) are available at MNRAS online.

**Table 1.** Program stars and observational parameters.
**Table 2.** Observing details.
**Table 3.** Median abundance errors.
**Table 4.** Stellar parameters.
**Table 5.** Chemical abundances.
**Table 6.** [C/N] abundances.
**Table 7.** [$\alpha$/Fe] abundances.
**Table 8.** [Sr/Ba] and [Ba/H] abundances.

**Table S1.** Fitting windows.
**Table S2.** Fitting windows for Na.
**Table S3.** NLTE corrections and continuum offsets for Ca II.
**Table S4.** CH evolutionary corrections.

Please note: Oxford University Press is not responsible for the content or functionality of any supporting materials supplied by the authors. Any queries (other than missing material) should be directed to the corresponding author for the article.

## APPENDIX A: CH FITS

The fits to the CH region across the wavelength region $4285 \leq \lambda \leq 4317$ Å are shown in Figs A1 and A2. Detections with best-fitting [C/Fe] value are shown in red, with the statistical fitting errors represented by the red shaded region. Those with non-detections are shown in blue, with the upper limit value being plotted. Regions shaded in grey were used in the $\chi^2$ calculations for determining detection limits.

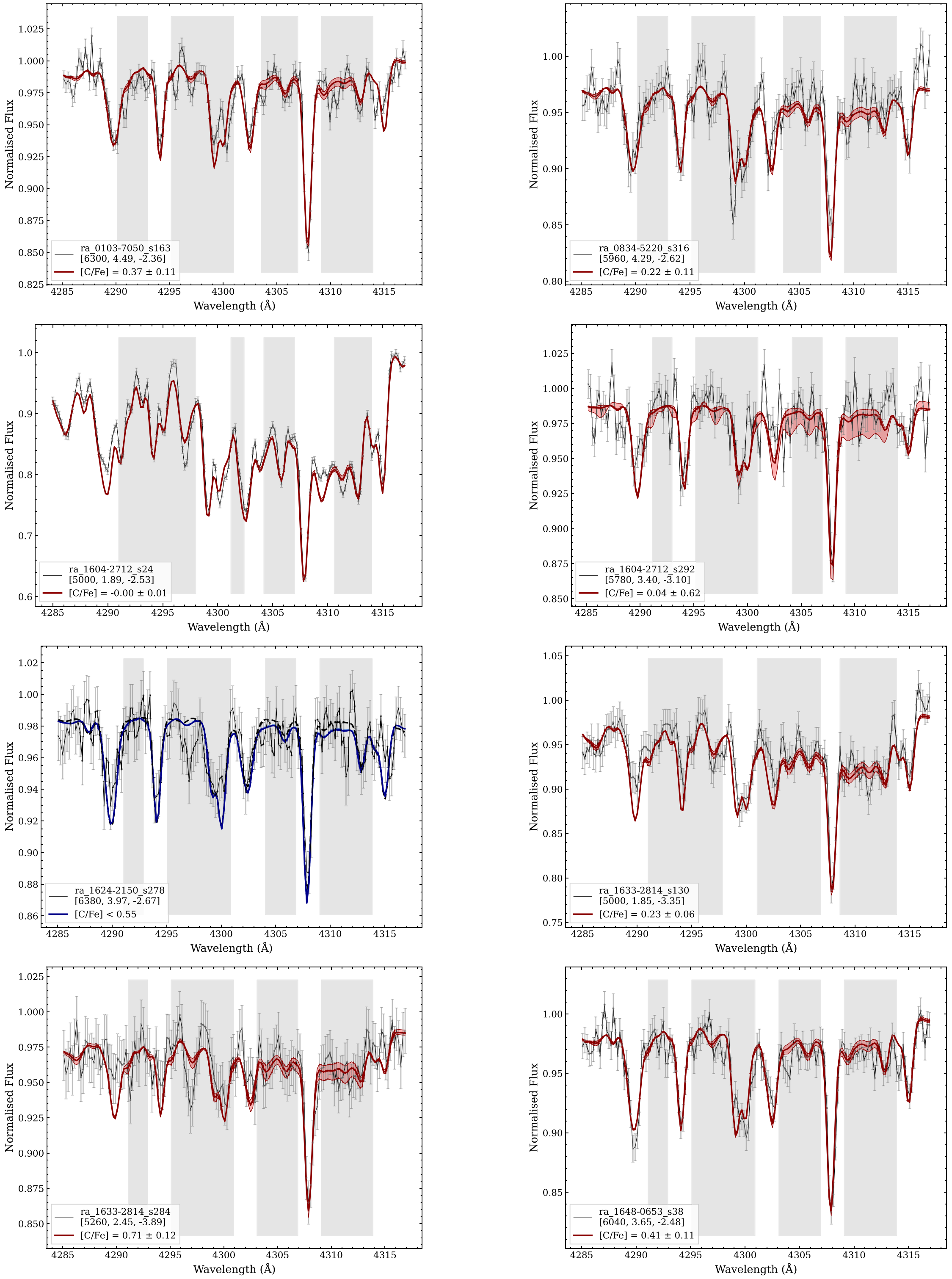

**Figure A1.** CH fits for the 16 sample stars across the wavelength region $4285 \le \lambda \le 4317$ Å. The observed data is shown by the thin line, and for detections: the thick line is best-fitting [C/Fe] value (alongside its fitting error; [C/Fe] value not corrected for evolutionary effects), with the statistical error shown by the shaded region. Those without detections have their upper-limit value fitted, shown by the thick line without shading. A reference synthetic spectrum with [C/Fe] = 0.5 is shown by the light dashed line, alongside a spectrum with [C/Fe] = −3.0 shown by the dark dashed line. Stellar parameters $T_{\rm eff}$, $\log g$, and [Fe/H] are found in the legend for each star. Grey shaded regions refer to regions used for $\chi^2$ calculations. Several prominent atomic lines are present, including Fe I at 4294.125 and 4307.901 Å, alongside Ca I at 4302.528 Å. Continues in Fig. A2.

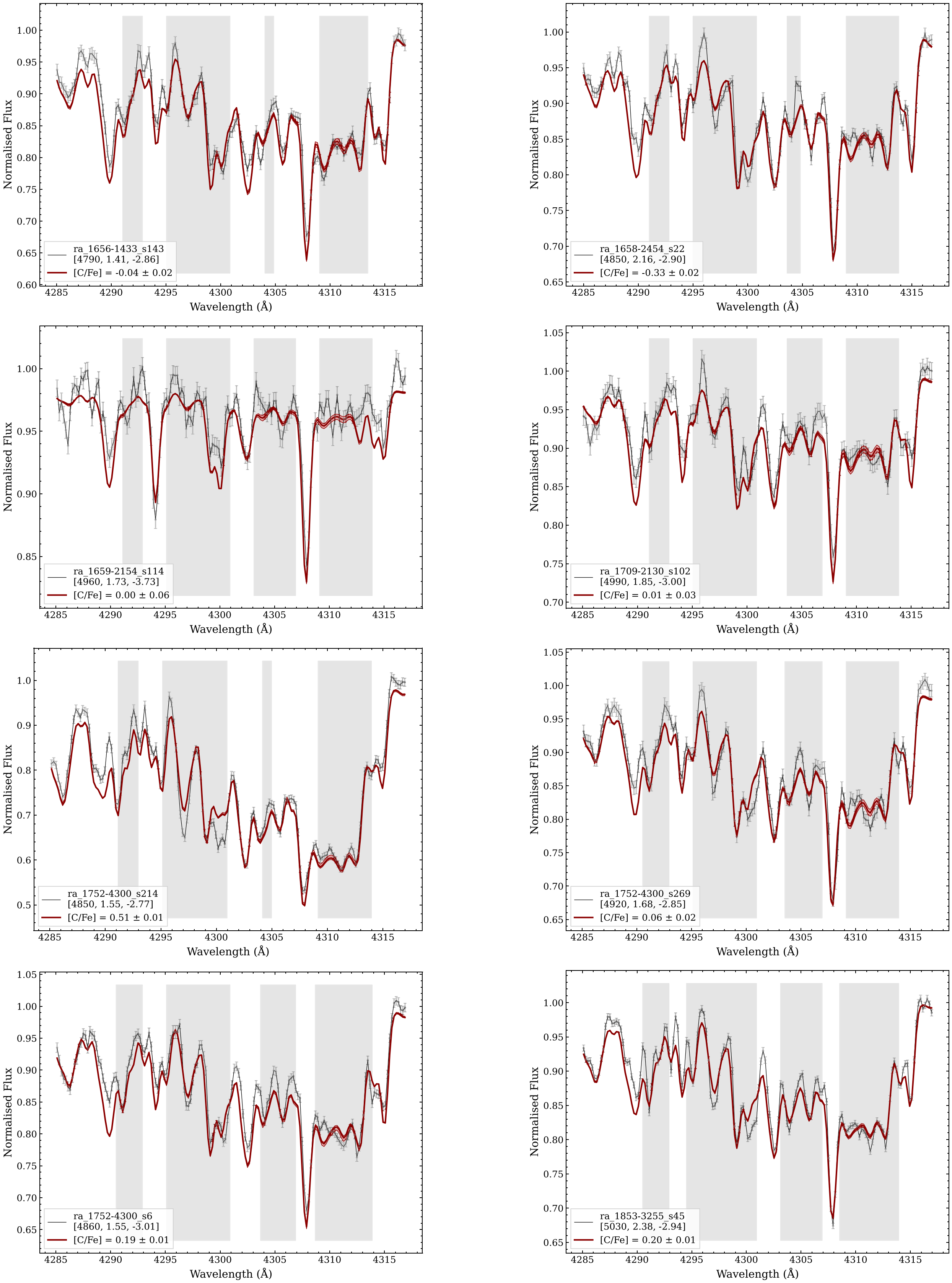

**Figure A2.** Continuation of Fig. A1.

## APPENDIX B: NH FITS

The fits to the NH region across the wavelength region $3355 \leq \lambda \leq 3365$ Å are shown in Figs B1 and B2. Plot format is identical to the CH fits described in Appendix A.

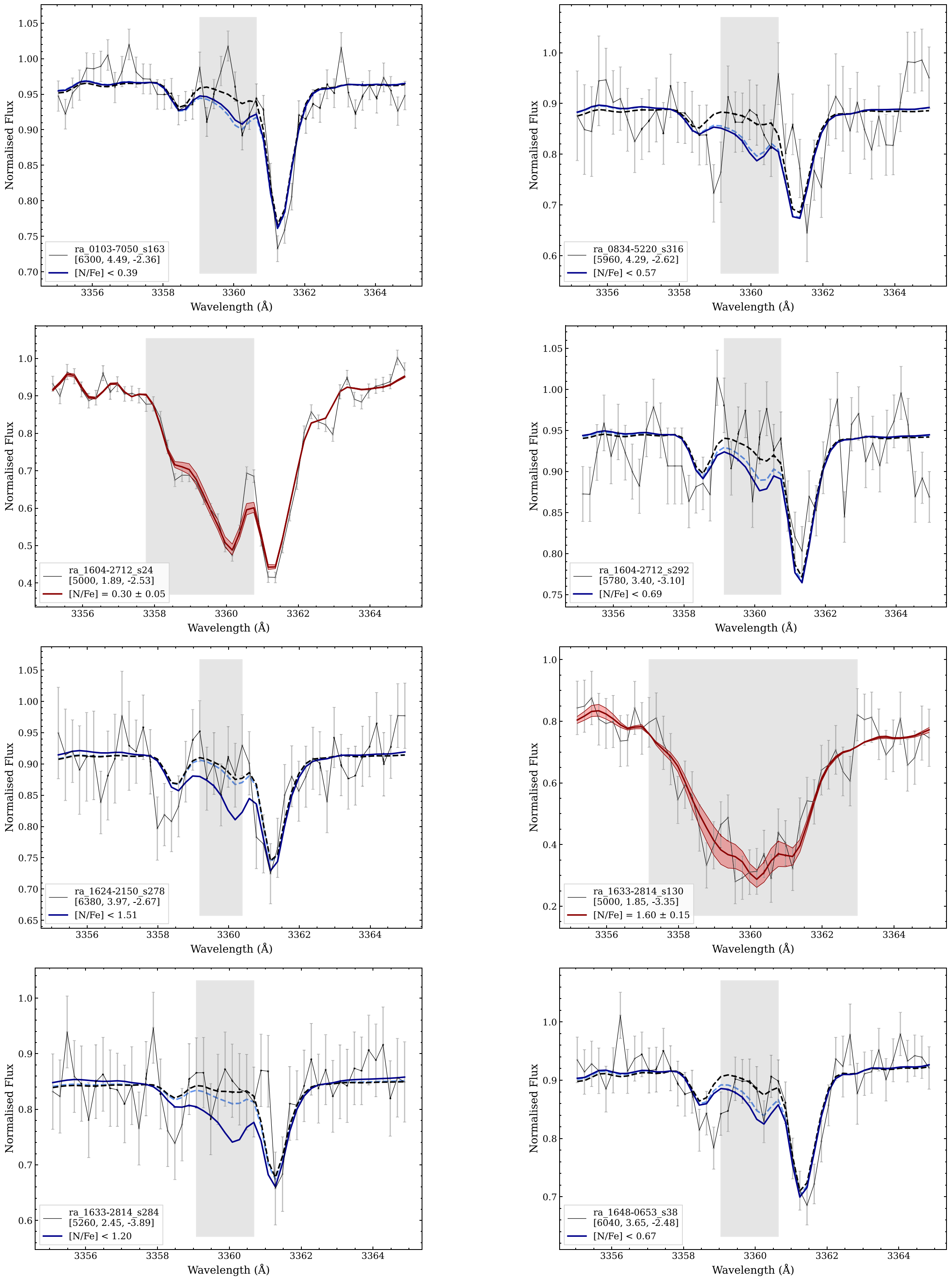

**Figure B1.** NH fits for the 16 sample stars across the wavelength region $3355 \leq \lambda \leq 3365$ Å. A reference synthetic spectrum with [N/Fe] = 0.5 is shown by the light dashed line, alongside a spectra at [N/Fe] = −3.0 shown by the dark dashed line. In our window, we have the atomic lines Cr I at 3358.491 Å and Ti II at 3361.212 Å present. Format is otherwise identical to the CH plots in Fig. A1. Continues in Fig. B2.

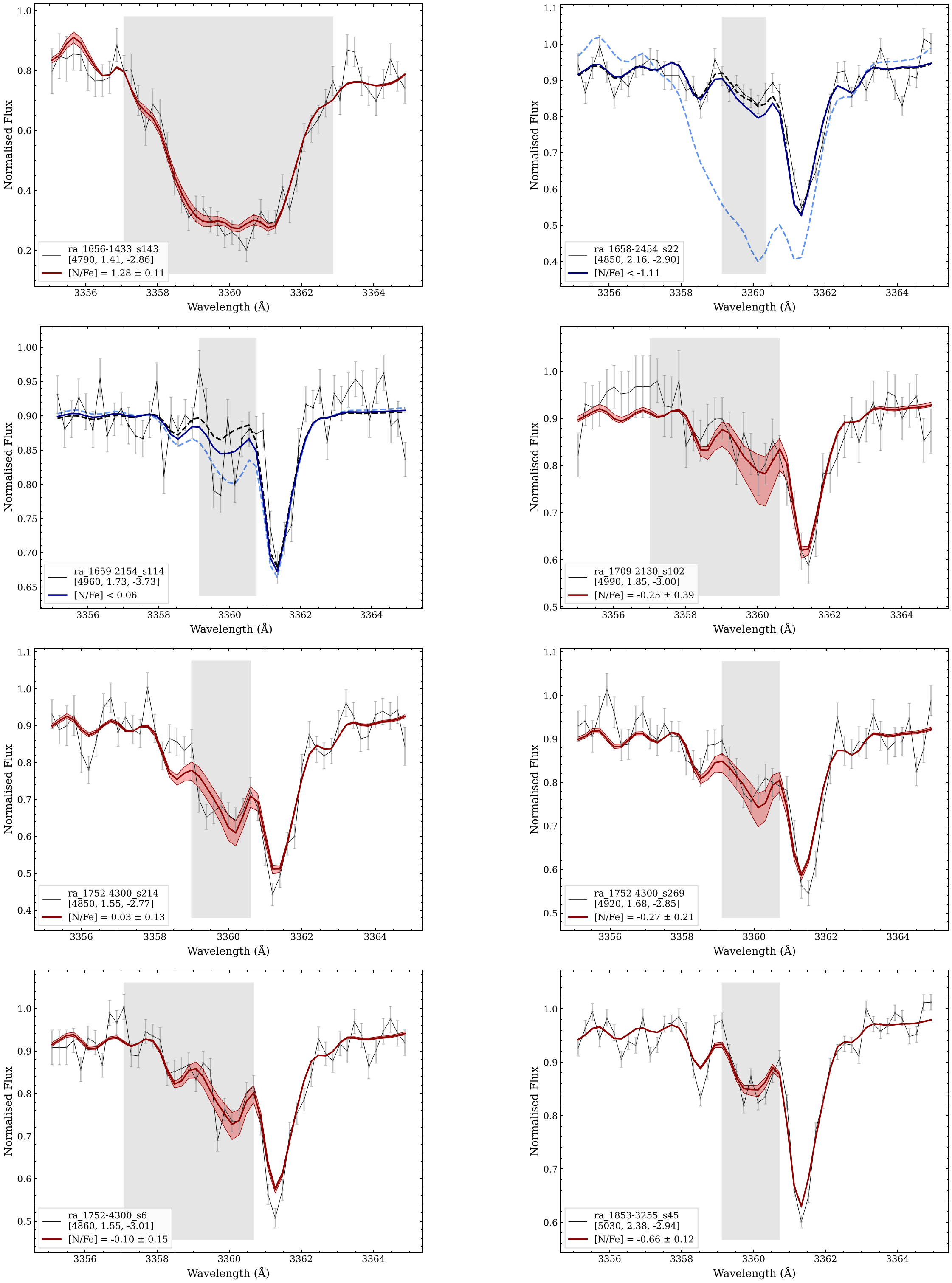

**Figure B2.** Continuation of Fig. B1.

This paper has been typeset from a TEX/LATEX file prepared by the author.